\documentclass[twocolumn]{aastex631}

\shorttitle{SN 1006 D6 Search}
\shortauthors{Shields et al.}
\graphicspath{{./}{figures/}}

\usepackage{graphicx}
\usepackage{subfigure}

\usepackage{savesym}
\savesymbol{tablenum}
\usepackage{siunitx}
\restoresymbol{SIX}{tablenum}

\begin{document}

\title{Searching for a Hypervelocity White Dwarf SN Ia Companion: \\ A Proper Motion Survey of SN 1006}

\author[0000-0002-1560-5286]{Joshua V. Shields}
\affiliation{Department of Physics and Astronomy, Michigan State University, East Lansing, MI 48824, USA}
\email{shield90@msu.edu}

\author[0000-0002-0479-7235]{Wolfgang Kerzendorf}
\affiliation{Department of Physics and Astronomy, Michigan State University, East Lansing, MI 48824, USA}
\affiliation{Department of Computational Mathematics, Science, and Engineering, Michigan State University, East Lansing, MI 48824, USA}

\author[0000-0003-2874-1196]{Matthew W. Hosek Jr.}
\affiliation{Department of Physics and Astronomy, University of California, Los Angeles, CA 90095, USA
}

\author[0000-0002-9632-6106]{Ken J. Shen}
\affiliation{Department of Astronomy and Theoretical Astrophysics Center,
University of California, Berkeley, CA 94720, USA
}

\author[0000-0002-4410-5387]{Armin Rest}
\affiliation{Space Telescope Science Institute, 3700 San Martin Drive, Baltimore, MD 21218, USA
}
\affiliation{Department of Physics and Astronomy, The Johns Hopkins University, 3400 North Charles Street,
Baltimore, MD 21218, USA
}

\author[0000-0001-9554-6062]{Tuan Do}
\affiliation{Department of Physics and Astronomy, University of California, Los Angeles, CA 90095, USA
}

\author[0000-0001-9611-0009]{Jessica R. Lu}
\affiliation{University of California, Berkeley, Department of Astronomy, Berkeley, CA 94720, USA
}

\author[0000-0001-7343-1678]{Andrew G. Fullard}
\affiliation{Department of Physics and Astronomy, Michigan State University, East Lansing, MI 48824, USA
}

\author[0000-0002-1652-420X]{Giovanni Strampelli}
\affiliation{Space Telescope Science Institute, 3700 San Martin Drive, Baltimore, MD 21218, USA
}

\author[0000-0001-6455-9135]{Alfredo Zenteno}
\affiliation{Cerro Tololo Inter-American Observatory, NSF’s National Optical-Infrared Astronomy Research Laboratory, 
Casilla 603, La Serena, Chile
}



\begin{abstract}

Type Ia Supernovae (SNe Ia) are securely understood to come from the thermonuclear explosion of a white dwarf as a result of binary interaction, but the nature of that binary interaction and the secondary object is uncertain.
Recently, a double white dwarf model known as the dynamically driven double-degenerate double-detonation (D6) model has become a promising explanation for these events. One realization of this scenario predicts that the companion may survive the explosion and reside within the remnant as a fast moving ($V_{peculiar} >1000$ km s$^{-1}$), overluminous ($L > 0.1 L_\odot$) white dwarf. Recently, three objects which appear to have these unusual properties have been discovered in the Gaia survey. 
We obtained photometric observations of the SN Ia remnant SN 1006 with the Dark Energy Camera over four years to attempt to discover a similar star. We present a deep, high precision astrometric proper motion survey of the interior stellar population of the remnant. We rule out the existence of a high proper motion object consistent with our tested realization of the D6 scenario ($V_{transverse} > 600$ km s$^{-1}$ with  $m_r < 21$ corresponding to an intrinsic luminosity of $L > 0.0176 L_\odot$). We conclude that such a star does not exist within the remnant, or is hidden from detection by either strong localized dust or the unlikely possibility of ejection from the binary system near parallel to the line of sight.

\end{abstract}


\section{Introduction} \label{sec:intro}

Type Ia supernovae (SNe Ia) are well-studied, highly energetic events that are fundamental drivers of galactic chemical enrichment \citep{timmes_galactic_1995, nomoto_nucleosynthesis_2013, kobayashi_new_2020} and that led to the discovery of the accelerating expansion of the Universe by allowing for secure measurements to distant galaxies \citep{riess_observational_1998, perlmutter_measurements_1999}. Despite the central role that these energetic events play in our understanding of the Universe and decades of focused research (e.g. see \citealt{maoz_observational_2014, ruiz-lapuente_surviving_2019}), we still do not know the progenitor system and explosion scenario that creates these events. SNe Ia arise from a carbon/oxygen white dwarf undergoing thermonuclear runaway \citep{pankey_possible_1962, colgate_early_1969}, but the circumstances that lead to this condition are uncertain. A misunderstanding of the underlying physics will result in uncertainties in our understanding of the Universe that is built upon these events.

SNe Ia progenitor scenarios are divided into two broad classes. In one major scenario, a white dwarf violently merges with a secondary white dwarf which leads to explosion \citep{iben_supernovae_1984, webbink_double_1984}. In the other, the primary white dwarf accretes material from a nearby secondary which also prompts thermonuclear runaway. This accretion scenario has many variations, with the secondary being either degenerate \citep{dan_prelude_2011} or non-degenerate \citep{whelan_binaries_1973, nomoto_accreting_1982-1, iben_interacting_1987, livio_progenitors_2000}. Significant work has been done attempting to disentangle progenitor scenarios and discover which, if any, of these processes are progenitors of SNe Ia, but finding strong support for any specific scenario has proven difficult (see \citealt{ruiz-lapuente_surviving_2019} for detailed discussion). One crucial, directly testable prediction comes from the secondary star in the binary system. In the violent merger scenario, the secondary is expected to be completely disrupted, while many accretion scenarios make the strong prediction that the secondary survives the explosion and exists within the resulting SN Ia remnant.

Identification of a surviving companion would lend powerful support to a corresponding accretion scenario based on the properties of the companion star \citep{marietta_type_2000, pakmor_impact_2008, shappee_type_2013, pan_impact_2012, pan_evolution_2013}. Galactic SN Ia remnants have been the subject of much scrutiny to discover a surviving companion, but no such companion has been unambiguously identified (e.g. see \citealt{ruiz-lapuente_binary_2004, ruiz-lapuente_no_2018, ruiz-lapuente_tychos_2019, ihara_searching_2007, hernandez_chemical_2009, kerzendorf_subaru_2009, kerzendorf_reconnaissance_2014, kerzendorf_tycho-b_2018, schaefer_absence_2012}). These works focused on identifying bright, non-degenerate companions tying back to the non-degenerate accretion scenario. However, mounting evidence including, but not limited to, non-detection of signatures of a non-degenerate companion in early \citep{hayden_rise_2010, bianco_constraining_2011, bloom_compact_2012, zheng_very_2013, olling_no_2015, marion_sn_2016, shappee_young_2016, shappee_strong_2018,  cartier_early_2017, miller_early_2018, holmbo_first_2019, fausnaugh_early-time_2021} as well as late times \citep{mattila_early_2005, leonard_constraining_2007, shappee_no_2013, lundqvist_hydrogen_2013, lundqvist_no_2015, sand_post-maximum_2016, sand_nebular_2018, graham_nebular-phase_2017, maguire_searching_2016, woods_no_2017, vallely_asassn-18tb_2019, tucker_nebular_2020}, disfavor the non-degenerate accretion scenario as an explanation for the bulk of SNe Ia, aligning with the non-detection of a non-degenerate surviving companion. Coincidentally, most surviving companion searches did not go deep enough to discover faint degenerate companions (e.g. white dwarfs) which have recently come to the forefront of the SNe Ia progenitor debate.

In this work, we test a specific realization of the Dynamically Driven Double-Degenerate Double-Detonation (D6) scenario \citep{guillochon_surface_2010, pakmor_sub-luminous_2010, pakmor_helium-ignited_2013, shen_ignition_2014}. In this scenario, the primary CO white dwarf undergoes unstable He accretion from a secondary degenerate He or CO white dwarf companion. The primary forms a thin He shell that detonates, which compresses the star and triggers thermonuclear runaway. If the He shell detonates early on in the accretion process, the secondary will survive the explosion and be flung out of the system with a minimum velocity of 1000 km s$^{-1}$ \citep{shen_three_2018}, significantly above that inherited by normal processes of stellar evolution barring specific dynamic interactions in the galactic center which are exceedingly rare \citep{hills_hyper-velocity_1988, brown_hypervelocity_2015, generozov_constraints_2022}. \citealt{shen_three_2018} discovered three hypervelocity white dwarfs in the field in the Gaia mission \citep{collaboration_gaia_2016, collaboration_gaia_2018} that lie in a peculiar region of the color luminosity diagram, aligning with this realization and providing the most powerful observational support that any progenitor scenario has seen. Combined with the mounting evidence against other established SNe Ia progenitor scenarios, this discovery suggests the possibility that most if not all normal SNe Ia arise from the D6 scenario. 

This scenario provides a testable hypothesis. If this realization of the D6 scenario is the generic explanation for SNe Ia, each SN Ia remnant must contain such a surviving companion. The SN ejecta that forms the remnant is ejected with a mean velocity $V_{ejecta, mean} \simeq$ \num[group-separator = \ ]{5000} km s$^{-1}$, and a maximum velocity $V_{ejecta, max} \geq \num[group-separator = \ ]{20000}$ km s$^{-1}$ (see e.g. \citealt{hillebrandt_type_2000}), far above the surviving companion's velocity. The ejecta slows upon colliding with the surrounding interstellar medium, but still leaves the companion contained by the SN remnant. We intended to test this realization of the D6 scenario by searching for a surviving D6 companion inside the SN 1006 remnant, which is uniquely suited for such a search. Galactic SN remnants generally trace the stellar population and therefore reside primarily within the Galactic plane (mean and standard deviation of Galactic remnant latitudes b = $0.117 \pm 2.787$ deg, \citealt{green_revised_2019}), which creates two significant problems: First, the Galactic plane is heavily obscured by dust and is prohibitively difficult to search for faint, blue objects (e.g. white dwarfs, see \citealt{ruiz-lapuente_no_2018}) with current observational constraints. Second, there is a high density of contaminating foreground and background interlopers which has two effects. The high density can contribute to source confusion and force the search to include sources which cannot be placed in front of or behind the remnant, obfuscating the search to the point of unfeasibility. SN 1006 uniquely resides nearby and high above the galactic plane with a galactic latitude b = 14.6 deg and a distance d = 2.17 $\pm$ 0.08 kpc \citep{winkler_sn_2002}\footnote{We note that there is some ambiguity on the distance to the remnant. \citealt{kerzendorf_search_2017} report the distance as 2.07 kpc $\pm 0.18$, but \citealt{winkler_sn_2002} report 2.17 $\pm 0.08$. We adopt this distance for the remainder of this work.}, leading to shallow foreground extinction ($A_v = 0.2154 \pm 0.0564$) and relatively little source confusion. Furthermore, a star moving 1000 km s$^{-1}$ in a transverse direction at such a close distance would show a very strong proper motion signal of 97.2 mas yr$^{-1}$, far above normal positional uncertainties of high precision astrometric measurements. These properties indicate that a high velocity D6 companion will be observable within the remnant if one exists. 

There have been multiple previous searches in SN 1006 for surviving companions, but they largely focused on discovering bright, non-degenerate donors and did not go deep or wide enough to discover a compact high velocity object in line with the predictions of the D6 scenario (e.g. \citealt{hernandez_no_2012} down to $m_r = 15$,  \citealt{kerzendorf_hunting_2012} down to $m_v = 19$ but a search radius r = 2 arcmin). Furthermore a D6 star in SN 1006 would have only had about $10^3$ years to evolve, orders of magnitude less time than the three candidates discovered in the field in \citealt{shen_three_2018} which are thought to be at least $10^5$ yr post explosion. \citealt{liu_observational_2021} showed that a D6 star will be significantly overluminous within $10^4$ years, but the appearance of a D6 star as young as in SN 1006 has not yet been observed, and an unusually high velocity remains the strongest signature of such an object. While \citealt{kerzendorf_search_2017} went down to $m_r = 21$  and directly sought to investigate the possibility of a white dwarf companion, a young D6 star's heavily uncertain appearance in color and luminosity could mean that it resides far off the standard white dwarf cooling track, which might have caused the star to elude this analysis as well.

In this work, we present a deep four year baseline astrometric proper motion survey of the stars inside SN 1006 using the Dark Energy Camera \citep{diehl_dark_2012, flaugher_dark_2015} to find the surviving companion predicted by the \citealt{shen_three_2018} realization of the D6 hypothesis. We measured and report the proper motions of over \num[group-separator = \ ]{2000} objects beyond the detection limit of Gaia. 

In Section \ref{sec:obs}, we present our observations and initial data reduction. In Section \ref{sec:methods}, we detail our astrometry and proper motion extraction. In Section \ref{sec:results}, we present the results of our survey and the constraints on a surviving companion in SN 1006 that follow. In Section \ref{sec:discussion} we consider confounding possibilities of non-detection and discuss high proper motion objects identified in our search. We conclude by summarizing our findings and discussing future work in Section \ref{sec:conclusions}.

\section{Observations \& Data Reduction} \label{sec:obs}
For our work, we acquired pre-existing photometry of the SN 1006 remnant from the nights of Jan 30 2017 and May 22 2018, and obtained new observations of the remnant on the night of Jan 22 2021. All data was captured using the Dark Energy Camera (\citealt{diehl_dark_2012}, \citealt{flaugher_dark_2015}) instrument mounted on the 4-m Blanco telescope located at the Cerro Tololo Inter-American Observatory (CTIO). Exposures were taken in 5 bands, but all data processing and analysis was performed on r-band observations to minimize atmospheric scattering as well as foreground dust extinction, allowing for higher astrometric accuracy and better measurements of faint, reddened sources. All r-band exposures were 50 seconds, stacked to create combined exposure times of 250 or 300 seconds depending on epoch. 

After standard calibration (bias correction, flat-fielding, and WCS) was done by the NSF NOIRLab DECam Community Pipeline \citep{valdes_decam_2014}, we reduced the data using the \texttt{Photpipe} pipeline as described in \cite{rest_testing_2005, rest_cosmological_2013}: Images were warped into a tangent plane of the sky using the ``SWarp'' routine \citep{bertin_terapix_2002}, before photometry of the stellar sources is obtained using the standard point spread function (PSF) fitting software DoPHOT \citep{schechter_dophot_1993}. We obtained observations of standard stars on the same nights as the photometric catalogues, which we used for calibration to obtain our photometric zeropoints.

Each epoch was comprised of either 5 or 6 dithered observations which were combined for each of the 62 individual CCDs. Multiple observations of the same star within one pixel coordinate (0.263 arcseconds) were matched and combined. Additional details about this initial matching are given in Appendix \ref{App:match}. Final stellar positions were then calculated using uncertainty weighted averages in both CCD pixel dimensions, and their uncertainties co-added using standard uncertainty propagation rules, decreasing uncertainties by a factor of $1/\sqrt{N}$. We note that, as dithering patterns did not observe all sources in each image, this factor was inconsistent depending on source position inside a CCD. With secure single epoch catalogue positions, we then needed to cross match sources across epochs to identify their movement and extract proper motions.

\section{Methodology \& Analysis} \label{sec:methods}

A source moving inside SN 1006 at $1000$ km s$^{-1}$, the minimum velocity in line with the predictions of the D6 scenario \citep{shen_three_2018}, would have a proper motion of $97$ mas yr$^{-1}$ assuming our chosen distance to the remnant of 2.17 kpc. We initially set out to discover any star within SN 1006 with a proper motion higher than 80 mas yr$^{-1}$ with no further restrictions other than being bright enough to be measurable in the DECam imagery. To recover proper motions with sufficient signal to noise to identify a surviving companion, we needed to discover a transformation from each individual instrumental CCD reference frame into one common reference frame. We chose to use the Gaia EDR3 catalogue \citep{gaia_collaboration_gaia_2021} to establish this common astrometric reference frame because it is currently the publicly available catalogue with the most precisely measured positions of the stars inside the remnant.  

\subsection{Building Our Proper Motion Catalogue}
We began by identifying a grid of 16 bright stars on each CCD with relatively small astrometric position uncertainties (see Equation \ref{eq:uncertainty} in Appendix \ref{App:pos}) that could be matched between our DECam and Gaia source catalogues. We used these as an initial guess for a second-degree polynomial (12 free parameters) that transforms from our DECam pixel coordinates to Gaia ICRS coordinates. Using this initial guess, we matched additional sources within one arcsecond (3.8 pixels) and one magnitude in the DECam and Gaia catalogues and refit the polynomial transformation. We note that Gaia G-band and DECam r-band are different filters, but we found that empirically the two bands are similar (r-band magnitude and G-band magnitude have a mean difference of 0.02 mag in our sample). The filters are also centered on similar wavelengths. Additionally, the magnitude matching was only used as a conservative safeguard against spurious matches. We finally performed a second iteration of this fitting process, matching the DECam catalogue to the Gaia catalogue with a polynomial up to fourth degree (30 free parameters) beginning with the previously matched stars as our initial guess to capture minor instrumental distortions. Between \num[group-separator = \ ]{4000} and \num[group-separator = \ ]{9000} stars were identified and matched between the Gaia and DECam catalogues in each CCD for this polynomial fitting step. Both iterations of the polynomial transformation were tested over a small range of polynomial orders to arrive at a transformation that produced strong agreement between Gaia and DECam positions. 

\begin{figure}
    \centering
    \includegraphics[width=.5\textwidth]{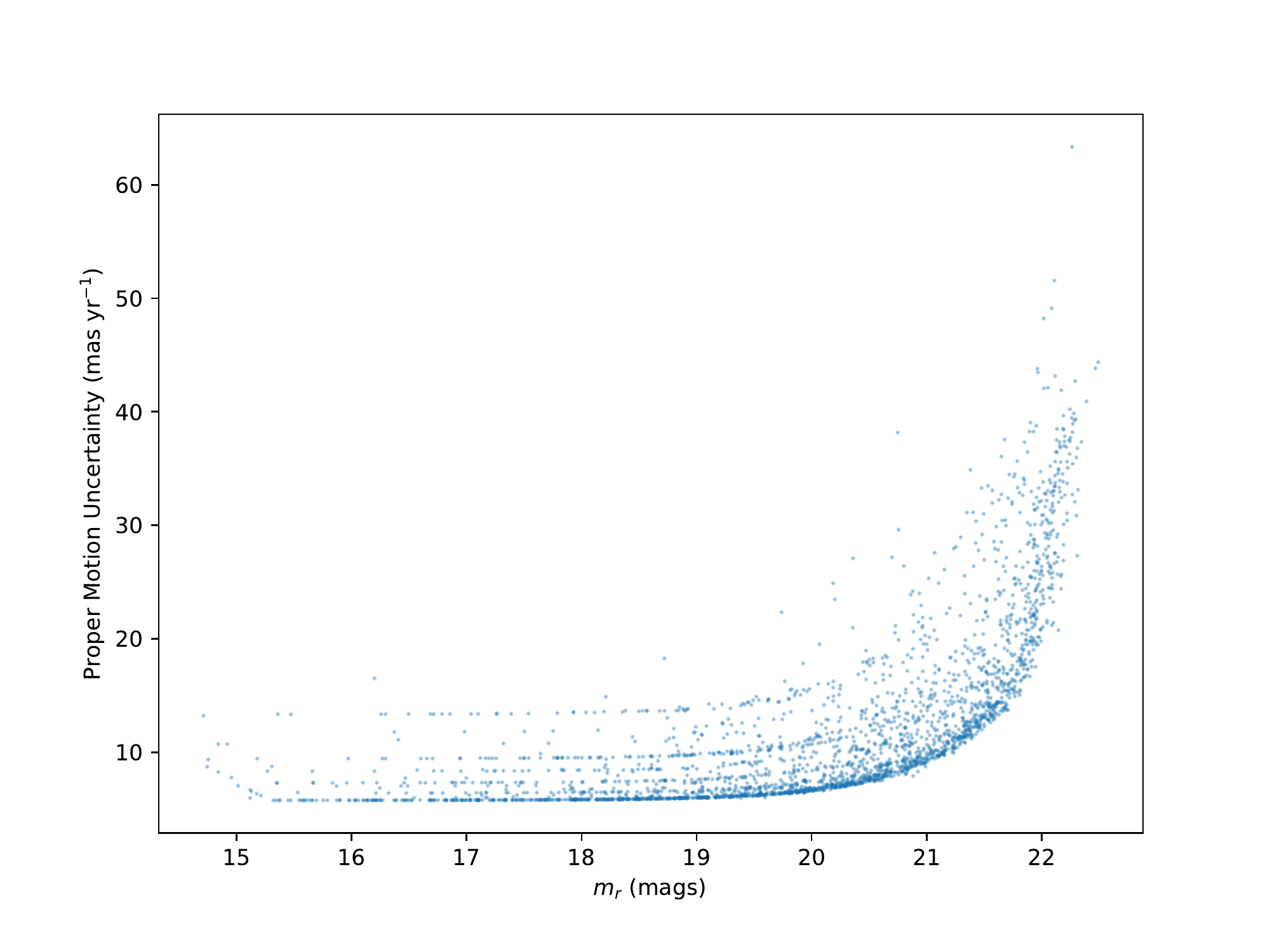}
    \caption{Proper motion uncertainty vs r-band magntitude. There are different systematic uncertainty floors depending on CCD dithering and number of observations for a source. The desired signal measurement was 80 mas yr$^{-1}$. 
}
    \label{fig:pm_unc}
\end{figure}

With this final polynomial transformation from the DECam instrumental reference frame to the Gaia ICRS defined reference frame, we matched DECam objects across all three epochs within one arcsecond and fit their motions independently in RA and Dec using $\chi^2$ minimization. Only stars detected in all three epochs were fit for proper motion. Final proper motion uncertainties are shown in Figure \ref{fig:pm_unc}. Uncertainties here are calculated using standard uncertainty propagation rules from $\chi^2$ minimization. The structure in the multiple systematic uncertainty floors seen in the figure traces back to stars at the edges of fields being observed in incomplete fractions of the imaging dithering patterns. For all but the faintest objects, our proper motions have uncertainties at least three times smaller than our desired signal of 80 mas yr$^{-1}$. We show additional independent verification of our proper motion measurements and comparisons to Gaia in Appendix \ref{App:compare}. Relative to Gaia, we find that over our whole sample, our proper motion measurements have a root mean square (RMS) difference average of 5.61 mas yr$^{-1}$ and RMS standard deviation of 5.14 mas yr$^{-1}$. This observed scatter is also much smaller than our desired signal of 80 mas yr$^{-1}$, allowing for proper motion measurements of sufficient quality to detect a surviving D6 companion within the remnant.

\begin{figure}
    \centering
    \includegraphics[width=.5\textwidth]{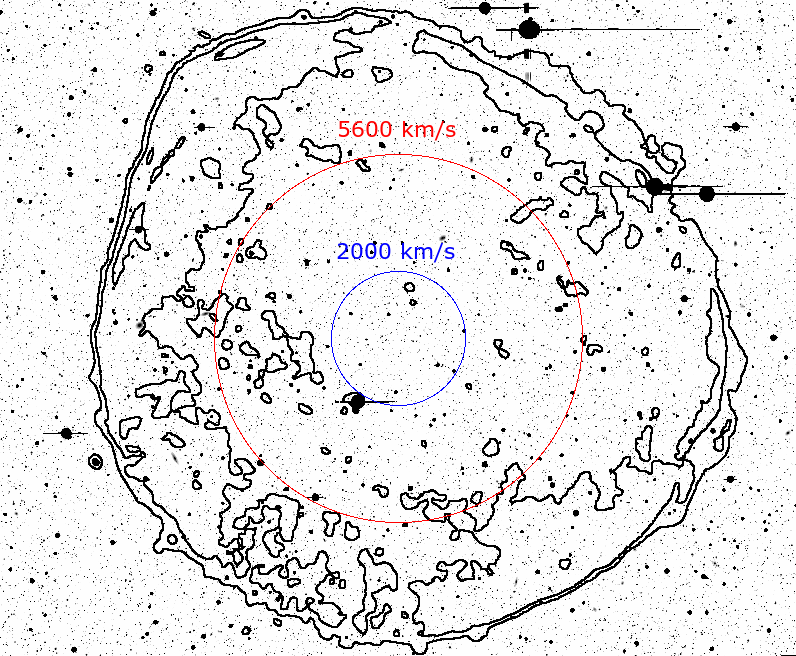}
    \caption{2017 DECam imagery of the SN 1006 remnant. The contours are CHANDRA X-ray data (0.5-0.9 keV) showing the position of the remnant. The circles indicate the search region (red) and the the likely maximum displacement of a D6 star (blue), as well as the physical transverse velocity of a star corresponding to the angular distance assuming a distance of the remnant. The larger search region allows for ambiguity on the center of the remnant.
}
    \label{fig:SN_1006}
\end{figure}

\subsection{Search Region and Parameter Restrictions}

The site of the SN Ia event that created SN 1006 is uncertain because density variations in the interstellar medium might have lead to a significant offset between the geometric center of the remnant and the site of explosion \citep{winkler_probing_2005, williams_azimuthal_2013}. We restricted our search to a 9 arcmin cone corresponding to a transverse velocity of $\sim 5600$ km s$^{-1}$ at a distance of 2.17 kpc, centered on the geometric center of SN 1006 reported as $15^h02^m55.4^s -41^{\circ}56'33''$ by \citealt{winkler_sn_2002}. This transverse velocity is far higher than the upper limit on the velocity expected for a surviving companion \citep{shen_wait_2017}, but the cone radius was chosen to allow for strong ambiguity on the site of the explosion.
Furthermore, because of this ambiguity, we made no directional proper motion cuts. This left us with \num[group-separator = \ ]{8123} stars for analysis. Our final catalogue can be found at 
\dataset[DOI: 10.5281/zenodo.6506198]{https://doi.org/10.5281/zenodo.6506198}
which includes \num[group-separator = \ ]{125116} sources both in and around the remnant, with sources delineated as either inside or outside the search region.

\section{Results} \label{sec:results}

\begin{figure}
    \centering
    \includegraphics[width=.5\textwidth]{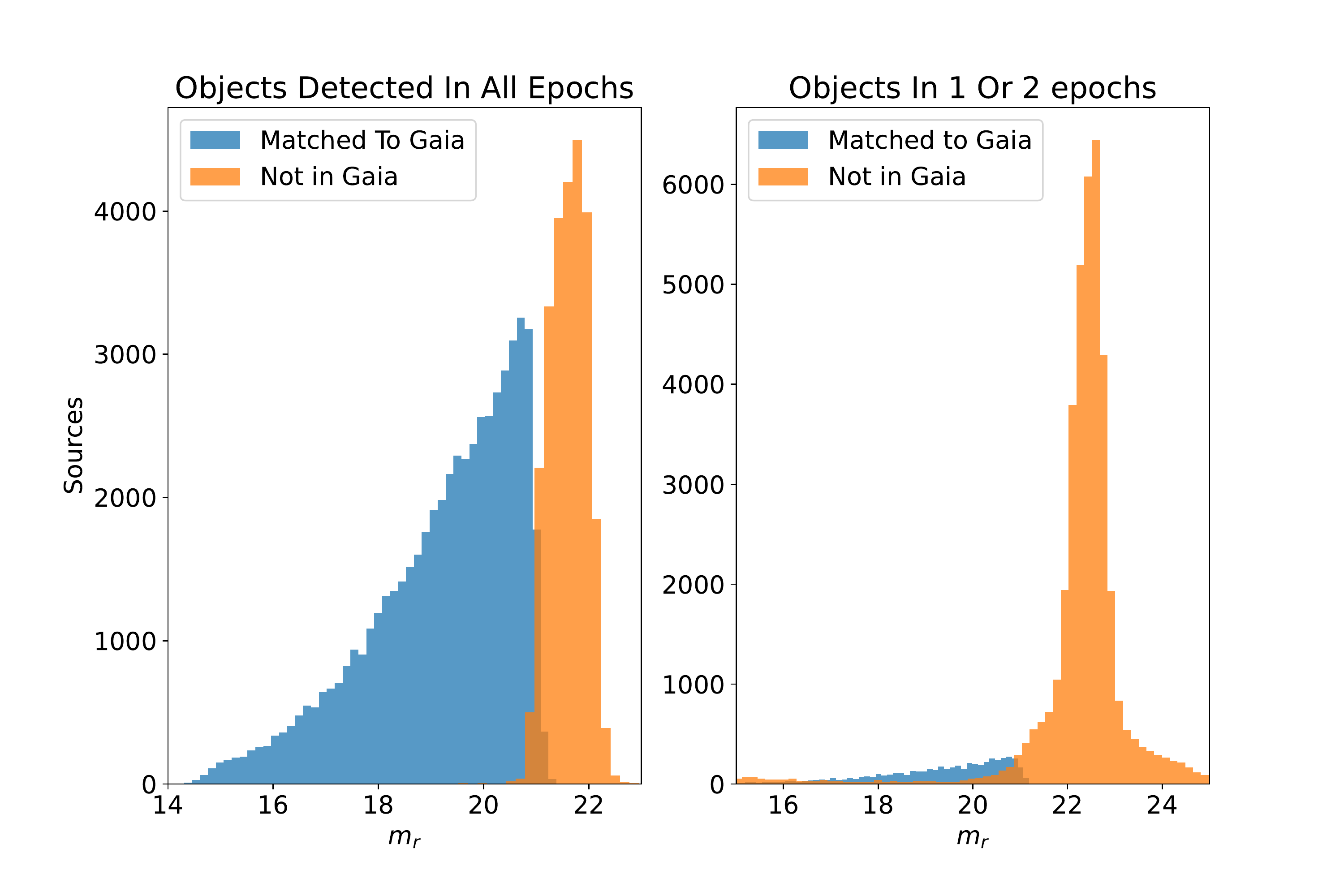}
    \caption{Histograms of all sources discovered in our search divided in to sources observed in all three epochs or in an incomplete fraction. We only extracted proper motions for objects discovered in all three epochs. 
}
    \label{fig:comp_hist}
\end{figure}

We conducted a high-precision proper motion astrometric survey of the stars within SN 1006 to search for a surviving companion predicted by the D6 scenario. In our sample, Gaia completeness appears to drop rapidly at $m_r = 21$ (see Figure \ref{fig:comp_hist}). Inside the remnant Gaia contains \num[group-separator = \ ]{5341} stars. We augmented the survey by measuring the proper motions of \num[group-separator = \ ]{2782} stars up to three magnitudes fainter than Gaia was able to detect. We comment on the high proper motion objects discovered at these faint magnitudes in Section \ref{sec:discussion}, but we chose $m_r = 21$ as the limiting magnitude for this work for three reasons: First, beyond this magnitude limit there are no available color or parallax measurements for our objects which makes it difficult to verify an object's identity as a surviving D6 companion. Second, completeness begins dropping quickly, which can be seen by the large number of objects detected in an incomplete fraction of our observations in Figure \ref{fig:comp_hist}. Third, proper motion uncertainties systematically grow beyond a third of our desired signal of 80 mas yr$^{-1}$ (see \ref{eq:uncertainty} in Appendix \ref{App:pos} and Figure \ref{fig:results}). 
 
Our limiting magnitude is multiple magnitudes fainter than a D6 star is expected to appear within the SN 1006 remnant. A D6 star similar to those discovered by \citealt{shen_three_2018} would appear between $17.5 < m_r < 19.2$. However, the three D6 candidates discovered in that work reside within the field, having likely had over $10^5$ years to radiate away excess energy injected from the explosion of the primary and reestablish equilibrium appearances. \citealt{liu_observational_2021} simulated the appearance of a D6 star shortly after explosion and found that the star would possess a luminosity of 10 $L_\odot$ ($m_r \approx$ 14.2 in SN 1006) $10^3$ years after the explosion, and remain above 1 $L_\odot$ ($m_r \approx$ 16.7 in SN 1006) for $10^7$ years. However, a D6 candidate as young as would appear in SN 1006 ($10^3$ years post SN) has not yet been observed. Thus, for our search, we conservatively considered all stars down to a magnitude of $m_r=21$. Additionally, while a D6 star inside the remnant is expected to possess a velocity of $>$1000 km s$^{-1}$, we initially considered down to a projected velocity equivalent of 800 km $s^{-1}$ to allow for a hidden radial velocity component and to compensate for measurement error. We then expanded our proper motion cut, detailed in the following paragraph.

\begin{figure}
    \centering
    \includegraphics[width=0.5\textwidth]{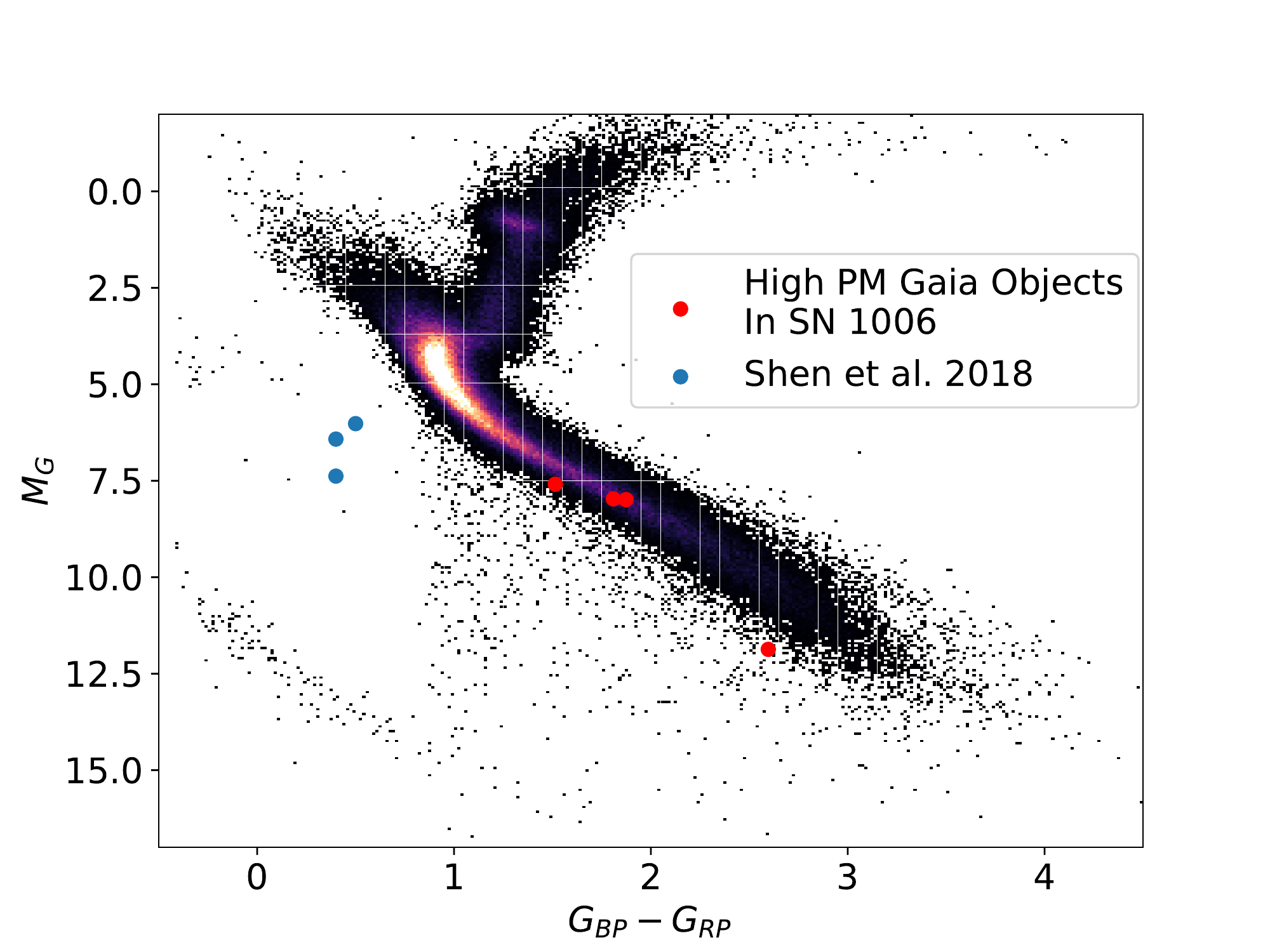}
    \caption{Color-Magnitude Diagram of \num[group-separator = \ ]{150000} secure parallax (parallax over error $> 30$), Gaia stars around SN 1006. The blue dots show the three D6 candidates discovered in the field by \citep{shen_three_2018} far off the main sequence. The red dots show the high proper motion ($> 500$ km s$^{-1}$) Gaia objects inside SN 1006 in our search. They lie on or close to the main sequence with ordinary colors. 
}
    \label{fig:weirdos}
\end{figure}

The results of our Gaia-DECam survey are shown in Figure \ref{fig:results}. We show the three previous D6 candidates from \citealt{shen_three_2018}, the expected parameter space of a similar star inside SN 1006, as well as our conservative limiting magnitude and proper motion cuts. We did not discover any high proper motion object in this space brighter than $m_r = 21$. To investigate the possibility of a high proper motion object slower than that expected \emph{a priori}, we investigated the 22 fastest of the 8123 stars within the remnant (the percentile equivalent of the 3-sigma highest proper motion outliers). Of these, 18 are too faint to be supported as the surviving companion ($m_r > 21$) without additional followup needed to rule each out from being a contaminating foreground star, a contaminating halo star, or from having a nonphysical proper motion measurement due to an undersampled (PSF) which leads to poor localization. The remaining four candidates are reported in Table \ref{tab:weirdos} under their Gaia identifiers and are marked in Figure \ref{fig:results}, reaching down to a proper motion of 62.5 mas yr$^{-1}$ or a projected velocity of 616 km s$^{-1}$. We additionally show a color-magnitude diagram of these four stars, the three D6 candidates from \citet{shen_three_2018}, as well as a sample of \num[group-separator = \ ]{150000} Gaia stars with secure parallax measurements (parallax\_over\_error $>$ 30) in and around the remnant in Figure \ref{fig:weirdos}. Unlike the three D6 candidates, the high proper motion Gaia stars in SN 1006 exist firmly on the main sequence and are each therefore unlikely to be the surviving companion. Additionally, Gaia EDR3 6004735811668137472 is the only object possessing a parallax that places it inside the remnant within uncertainties, while the other three have parallax measurements that point to foreground star identifications.

\begin{table*}[]
    \centering
    \begin{tabular}{c|c|c|c|c|c|c}
         GAIA Source ID & $m_G$ & $G_{BP} - G_{RP}$ & $m_r$ & Parallax & Proper Motion  & Projected Velocity Inside Rem\\
          &  &  &  & (mas) & (mas/yr) & (km/s) \\
          \hline
         6004785431417429120& 15.35 & 1.87 &  15.35 &$ 3.37\pm 0.04$ &$ 67.4\pm 0.1$ &  $693.3\pm 1.0$\\ 
         6004784984740826752& 20.20 & 2.60 &  20.20 &$ 2.16\pm 0.81$ &$ 82.2\pm 1.4$ &  $845.6\pm 14.4$\\
         6004735811668137472& 18.64 & 1.51 &  18.64 &$ 0.62\pm 0.27$ &$ 44.2\pm 0.4$ &  $454.7\pm 4.1$\\ 
         6004735094407417344& 15.26 & 1.81 &  15.26 &$ 3.48\pm 0.04$ &$ 60.2\pm 0.1$ &  $619.3\pm 1.0$\\ 
         
    \end{tabular}
    \caption{High proper motion Gaia sources in SN 1006. Projected velocity assumes a distance of 2.17 kpc and not a distance implied by the parallax measurement in case of a spurious parallax measurement that would cause us to miss a surviving companion. Proper motions shown here Gaia reported measurements.}
    \label{tab:weirdos}
\end{table*}

\subsection{Constraints On Intrinsic Stellar Luminosity}

\begin{figure}[]
    \centering
    \includegraphics[width=0.4\textwidth]{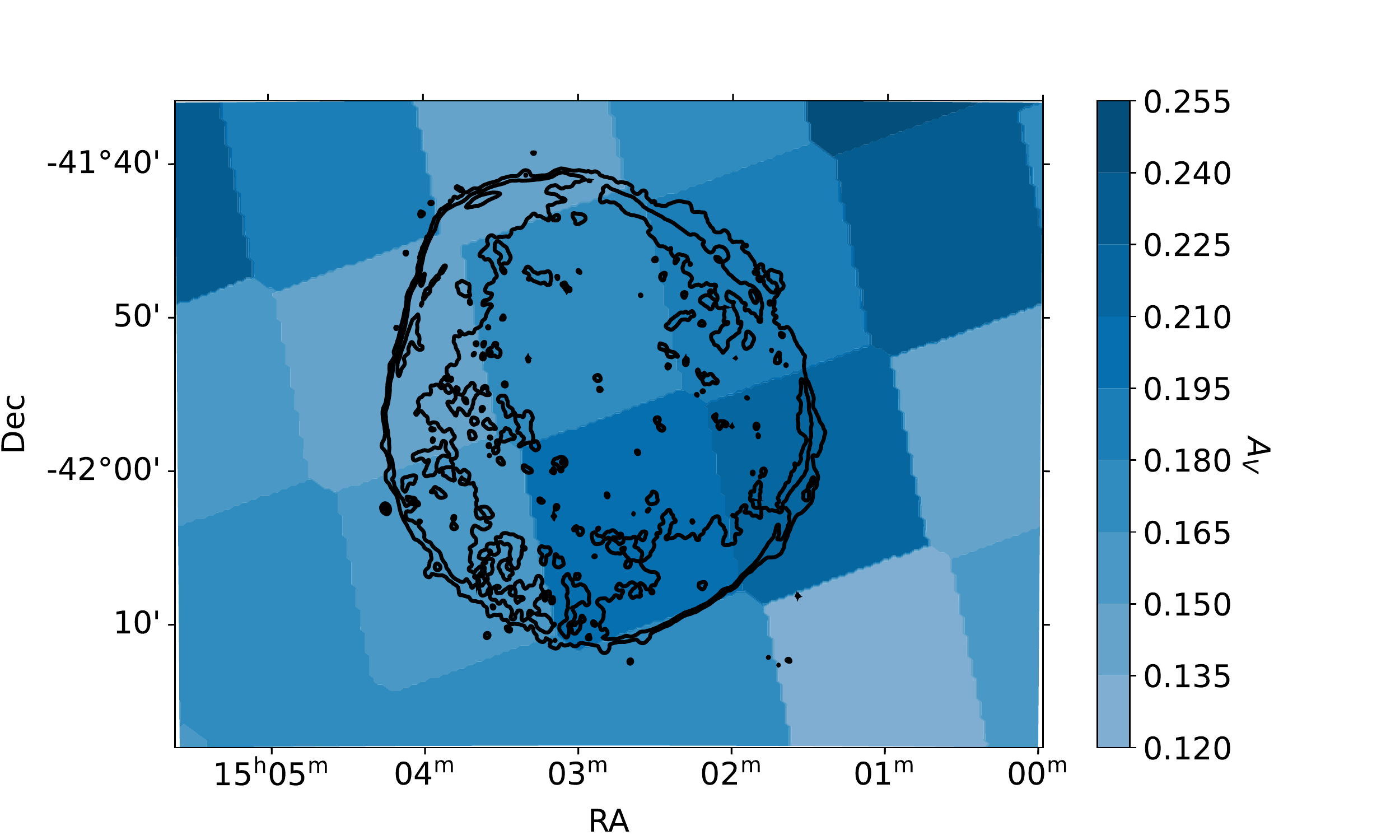}
    \includegraphics[width=0.4\textwidth]{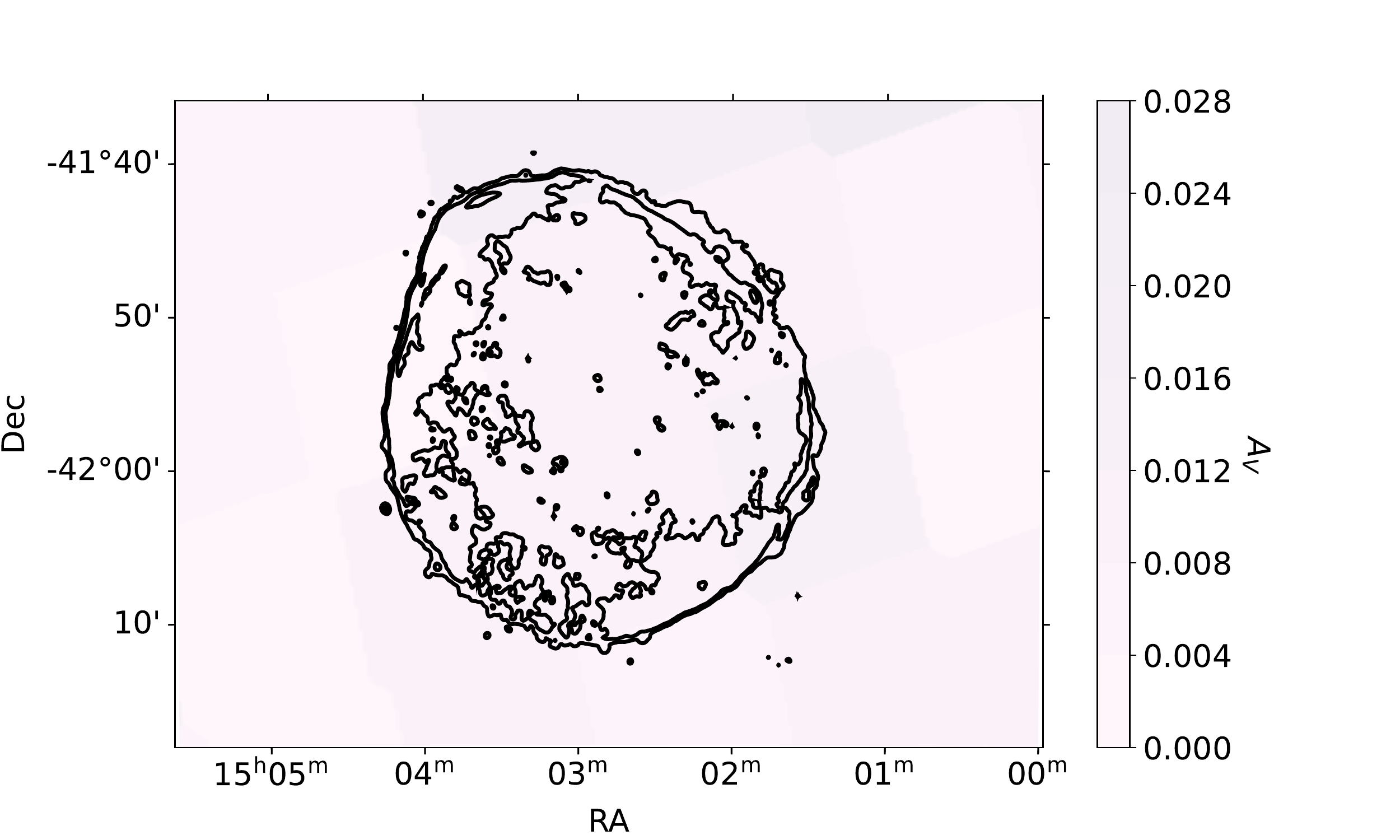}
    \caption{\citealt{guo_three-dimensional_2021} three-dimensional dustmaps sampled in the direction of the SN 1006 remnant with x-ray contours displayed to show the position of the remnant. The top shows the maps sampled at the distance of the remnant (2.17 kpc). The bottom shows the difference between the absorption sampled at 1.94 and 2.41 kpc ($2.17 \pm 0.24$, 3 standard deviations of the uncertainty on the distance, 0.08, respectively). Both maps are placed on a consistent color-scale to emphasize the lack of additional dust extinction measured within the remnant. $A_V$ is calculated assuming $E_{B-V} = 3.1A_V$ following \citealt{schlafly_measuring_2011}.
}
    \label{fig:dust_map}
\end{figure}

We investigated the intrinsic luminosity constraints of our survey. To estimate the foreground extinction between us and the SN 1006 remnant, we used the \cite{guo_three-dimensional_2021} southern sky three-dimensional dust maps, shown in Fig \ref{fig:dust_map}. The map queried at the distance of the remnant gives an extinction $E_{B-V} = 0.0673 \pm 0.0176$.  Assuming an $R_V = 3.1$ F99 reddening law following \citealt{schlafly_measuring_2011}, we calculate $A_V = 0.2154 \pm 0.0564$.  Furthermore, using a resulting ${A_r}/{A_V} = 0.89$, we estimate $A_r = 0.192 \pm 0.050$. Adopting $A_r = 0.192$, a distance modulus of 11.68 corresponding to a distance of 2.17 kpc, and a bolometric magnitude equal to r-band, a $m_{r}=21$ object posses an intrinsic luminosity of $L = 0.0176 L_\odot$. We did not detect a high proper motion object with unusual colors brighter than this luminosity within SN 1006.

\begin{figure*}
    \centering
    \includegraphics[width=1\textwidth]{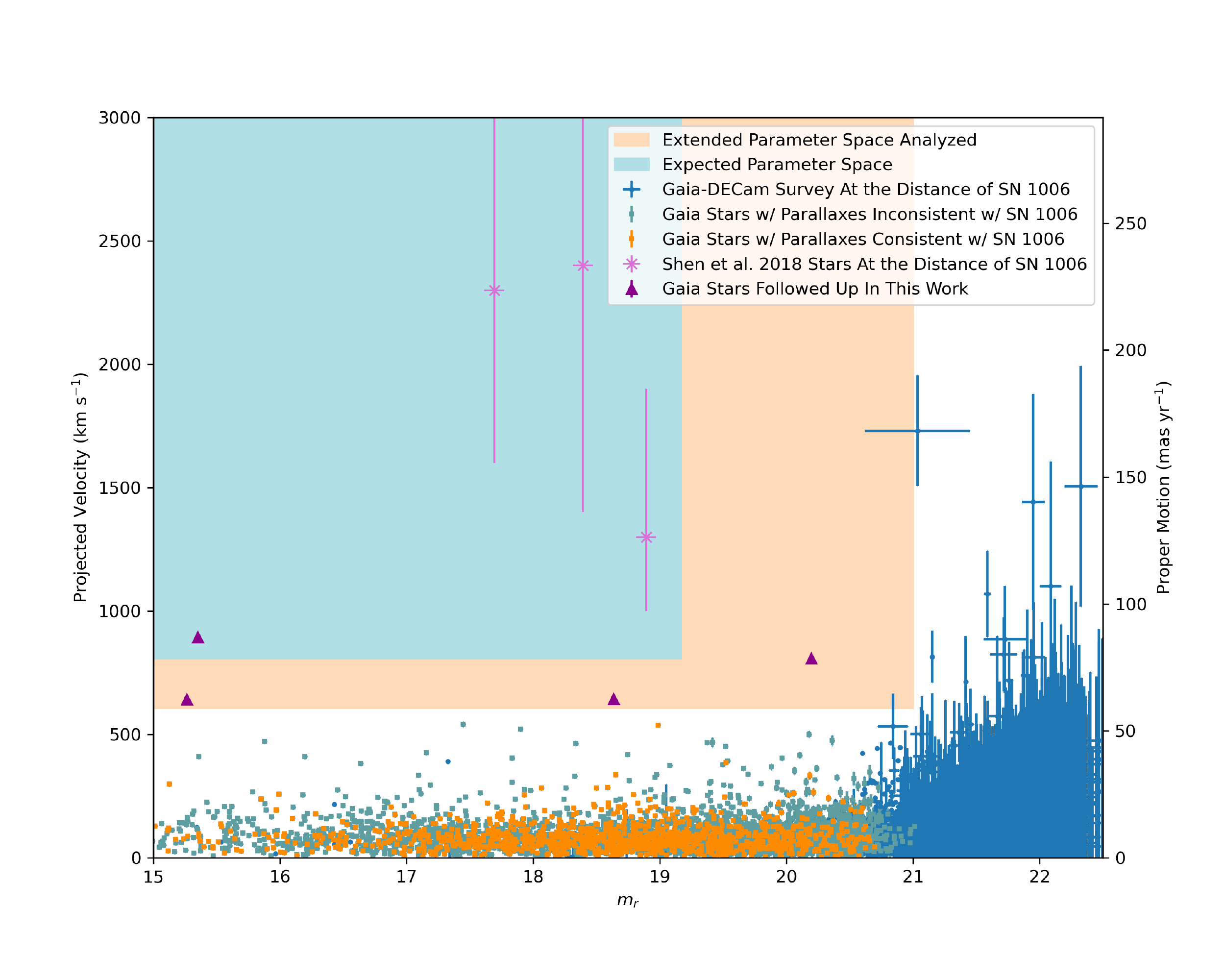}
    \caption{Results of proper motion survey showing apparent r-band magnitude against proper motion measurement and calculated transverse velocities assuming a distance of 2.17 kpc for DECam sources. The three \citealt{shen_three_2018} stars have corrected apparent magnitudes as they would appear at the same distance with uncertainties and including foreground extinction. A surviving white dwarf companion in accordance with the predictions of the D6 scenario was expected to lie in the shaded region with the previously discovered D6 stars. The four Gaia stars in the analyzed region are shown in Table \ref{tab:weirdos} and Figure \ref{fig:weirdos}, and are discussed in Section \ref{sec:results} along with the high proper motion objects fainter than 21. 
}
    \label{fig:results}
\end{figure*}

\section{Discussion} \label{sec:discussion}

We have investigated a prediction of one realization of the D6 scenario which might be a generic explanation for SNe Ia. We have no observationally motivated constraints for how a D6 star would appear shortly after the explosion of the primary. The three candidates presented in \citealt{shen_three_2018} were discovered in the field of the galaxy, and are estimated to be older than $\sim10^5$ years post explosion. Theory suggests that a young D6 star would be brighter than $m_r > 19$, but a significant inherited velocity remains the strongest observable property of such a star. We have investigated the stars in SN 1006 for this signature. Our survey places strict limits on the parameter space that a surviving D6 companion could exist inside SN 1006. Previous direct searches went down to $m_r = 15$ \citep{hernandez_no_2012} and $m_V = 19$ \citep{kerzendorf_hunting_2012}. \citealt{kerzendorf_search_2017} went down to $m_r = 21$ but only sought to investigate objects closely following the white dwarf cooling track. Here, we rule out the possibility of a high velocity surviving companion in the remnant down to $m_r = 21$.

\subsection{Confounding Possibilities}
We did not detect an overluminous surviving white dwarf companion with a high enough proper motion to be consistent with the D6 scenario inside SN 1006. We explored possibilities for a surviving D6 companion to have gone undetected in our analysis. Our search targeted stars with large transverse velocities detectable through proper motion measurements. As a result, we identified two following outstanding reasons that a D6 star in the remnant might have gone undetected in our study: First, the star may have inherited an exceedingly large radial velocity with a small transverse velocity by being ejected near parallel to our line of sight. Second, the star may be hidden by significant, unforeseen dust obscuration prompted by interaction with the SN.

\begin{figure}
    \centering
    \includegraphics[width=.5\textwidth]{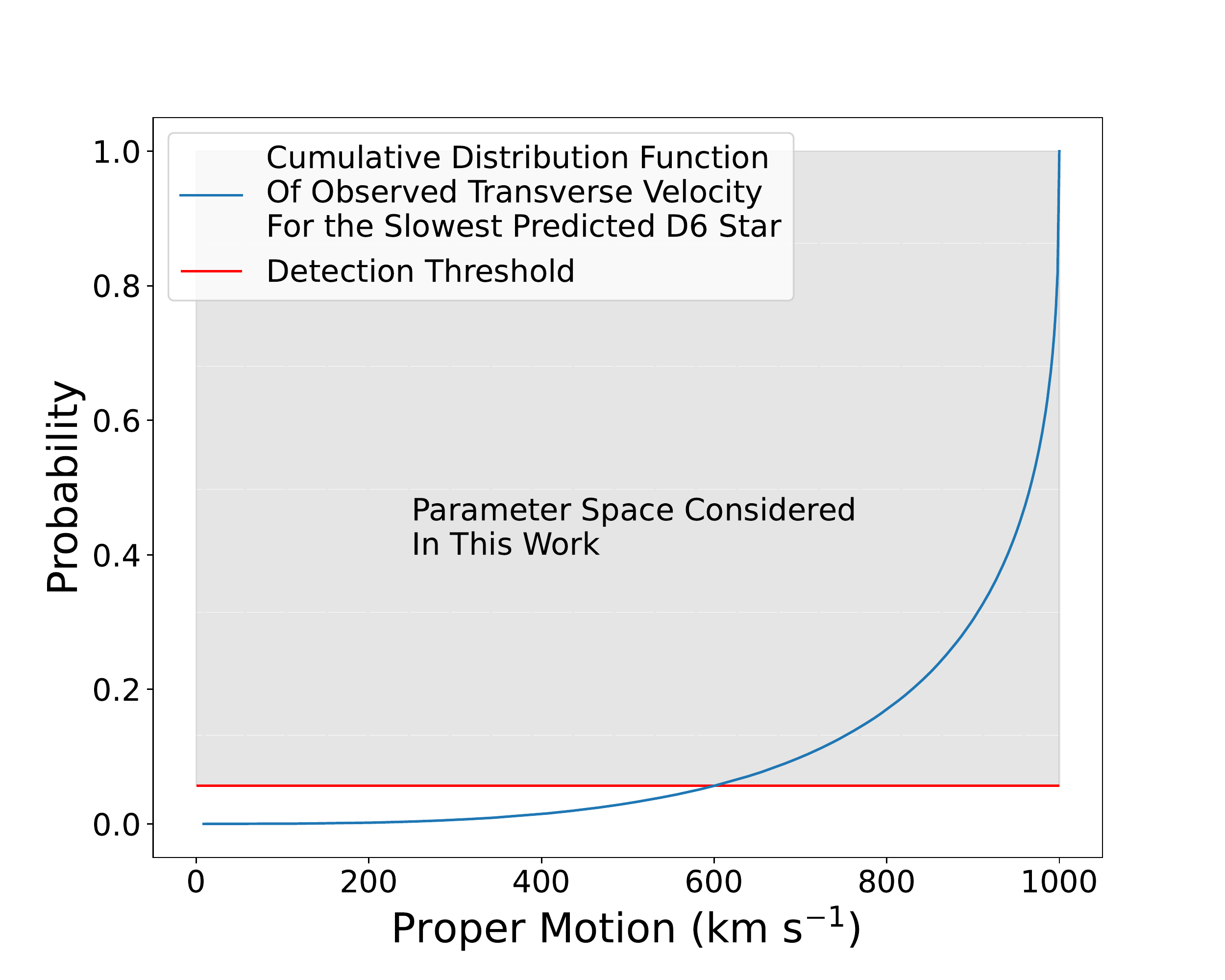}
    \caption{Results of our MC simulation showing the observed transverse velocity of a 1000 km s$^{-1}$ star, the minimum expected velocity of a D6 star, traveling in a random direction. We examined all stars with proper motions corresponding to transverse velocities of above 600 km s$^{-1}$ or proper motions of 58.3 mas yr$^{-1}$ (the top 22 proper motion objects in our survey). As shown, this corresponds to examining 94.3\% of this distribution. A faster velocity quickly shrinks the probability space, effectively moving further down the shallow tail of the distribution. 
}
    \label{fig:direction}
\end{figure}

We explore the possibility that a star launched with a significant velocity in a random direction did not inherit a large transverse velocity, with the majority of the velocity hidden in the radial direction. We performed a Monte Carlo simulation of a star launched in a random direction with a velocity of 1000 km s$^{-1}$, the lowest theoretically predicted velocity of a D6 star. The resulting probability distribution of the observed tangential velocity measured for such a star is presented in Figure \ref{fig:direction}. Our experiment and analysis would have detected the star above 94.3\% of the time. However, a D6 star likely inherits a velocity far greater than 1000 km s$^{-1}$, quickly shrinking the unexamined portion of this distribution (e.g., a 1200 km s$^{-1}$ star would have been discovered 97.3\% of the time). This parameter space could be examined with detailed radial velocity measurements of the stars in SN 1006, but remains a small, outstanding possibility. 

We also consider the possibility that a D6 star could have been heavily enshrouded by dust from the remnant and thus appeared fainter than 21st magnitude in r-band. In Figure \ref{fig:dust_map}, we show the V-band absorption enclosed by the remnant, approximated by sampling the \citet{guo_three-dimensional_2021} dust maps in front of and behind three times the uncertainty on the distance to the remnant. We see no evidence of additional dust absorption in the remnant on large enough scales to be detectable by these dust maps, with an angular resolution of 13.7 arcmin. Any obscuring dust capable of hiding the surviving companion would need to be localized around the star itself, such as dust produced by a strong stellar wind prompted by interaction with the SN explosion.

\subsection{Faint High Proper Motion Outliers}

As a result of this survey, we discovered six faint, high proper motion objects between 21 and 22.5 mag listed in Table \ref{tab:cands} and seen in the faint end of Figure \ref{fig:results}. These objects are too faint to have color information, and we note that these objects have large proper motion uncertainties due to their poorly constrained initial positions tracing back to their faint appearances, and thus the objects are mostly less than one or two sigma above our velocity threshold. While one of these objects may be the surviving companion in line with the significant dust obscuration scenario detailed above, we note that each is also likely a result of statistical sampling uncertainties and the large number of sources examined. Furthermore, we examined the Spitzer Enhanced Imaging Products \citep{https://doi.org/10.26131/irsa3, fazio_infrared_2004, werner_spitzer_2004} combined 3.6, 4.5, 5.8, and 8.0 micron image catalogue for coincident infrared sources with our high proper motion objects and did not discover any aligned sources. This provides strong evidence against dust obscuration, as absorbed light should be re-radiated in the infrared and make the source easily detectable. We however note that spectral followup of these objects would concretely support or reject their respective statuses as the surviving companion by providing both color information that could place them on the main sequence or in an unusual portion of the color magnitude diagram, as well as radial velocity measurements that could support or oppose high intrinsic velocities as opposed to modest intrinsic velocities that appear large due to foreground nature.

\begin{table*}[]
    \centering
    \begin{tabular}{c|c|c|c|c|c}
         Target & RA & Dec & $m_r$ & Proper Motion  & Projected Velocity\\
          & (deg) & (deg) &  & (mas/yr) & (km/s) \\
          \hline
         Candidate 1& 225.618762 & -41.856965 &  22.33 &$ 146.4\pm 47.3$ &$ 1506.0\pm 486.6$  \\
         Candidate 2& 225.801826 & -42.041687 &  22.09 &$ 107.0\pm 49.1$ &$ 1100.7\pm 505.1$ \\ 
         Candidate 3& 225.823198 & -42.064307 &  21.73 &$ 86.0 \pm 21.0$ &$ 884.7\pm 216.0$ \\
         Candidate 4& 225.697236 & -41.985479 &  21.72 &$ 80.2 \pm 14.6$ &$ 825.0\pm 150.2$ \\
         Candidate 5& 225.916877 & -41.995497 &  21.59 &$ 103.9\pm 17.1$ &$ 1068.8\pm 175.9$ \\
         Candidate 6& 225.586830 & -41.888318 &  21.04 &$ 168.3\pm 21.8$ &$ 1731.0\pm 224.1$ \\
    \end{tabular}
    \caption{Discovered faint high proper motion objects in SN 1006. Projected velocity assumes a distance of 2.17 kpc. These objects are too faint to be supported as D6 candidates in this work but remain interesting candidates. Additionally, we investigated candidate 6 and found it to have an unreliable measurement due to its psf overlapping with a nearby bright star.}
    \label{tab:cands}
\end{table*}

\section{Conclusions and Future Work} \label{sec:conclusions}

We present a deep four year baseline astrometric survey of the stars in the SN 1006 remnant. We do not detect an overluminous, high proper motion white dwarf similar to that predicted by a realization of the D6 scenario presented in \citealt{shen_three_2018}. This result suggests that this realization of the D6 scenario is not generically responsible for SNe Ia. Alternatively, this result might be in line with the recent result of \citealt{pakmor_fate_2022}, which showed that a SN Ia like-event can be created through the detonation of both the primary and the secondary white dwarf in the D6 scenario. 

We investigated possibilities that a surviving companion star similar to or brighter than those detected in the field could have gone undetected in our study. 
We find that: \begin{itemize}
    \item There is less than a 5.7\% chance that a star with an intrinsic velocity of 1000 km s$^{-1}$ could have been ejected from the system near parallel to our line of sight, and have inherited a small enough proper motion to have remained undetected. This possibility shrinks quickly as the velocity of the surviving companion increases.
    \item There is no significant additional large scale dust created by or coincident with the SN remnant. Dust capable of obscuring the surviving companion would need to be on angular scales significantly smaller than 13.7 arcmin or a physical scale of 8.6 pc at the distance of the remnant, localized around the surviving companion itself. 
\end{itemize}

We briefly consider the possibility of detecting a surviving D6 companion with an assumed luminosity of about 0.1 $L_\odot$ in other Galactic SN Ia remnants. We did not detect an unambiguous, overluminous white dwarf companion inside SN 1006 as predicted by our tested realization of the D6 theory, however there exist tens of SN Ia remnants in the Galaxy where a similar study could potentially be performed to search for a similar star. Upon inspection, we find that only three unambiguously classified SN Ia remnants exist with comparable distances to SN 1006, where a D6 star is approaching detection limits. These three remnants are RCW 86, Tycho's SNR, and G272.2-3.2. These remnants all lie within or close to the Galactic plane behind 4.4, 3.5, and 2.6 magnitudes of V-band extinction respectively \citep{schlafly_measuring_2011}. This poses a challenging obstacle for repeating a high precision proper motion survey within these remnants, where a 0.1 $L_\odot$ star could be as faint as $m_r \sim 24$.

\section{Acknowledgments}

This work has made use of data from the European Space Agency (ESA) mission
{\it Gaia} (\url{https://www.cosmos.esa.int/gaia}), processed by the {\it Gaia}
Data Processing and Analysis Consortium (DPAC,
\url{https://www.cosmos.esa.int/web/gaia/dpac/consortium}). Funding for the DPAC
has been provided by national institutions, in particular the institutions
participating in the {\it Gaia} Multilateral Agreement.

This work is based [in part] on observations made with the Spitzer Space Telescope, which is operated by the Jet Propulsion Laboratory, California Institute of Technology under a contract with NASA.

\software{Astropy \citep{the_astropy_collaboration_astropy_2013, collaboration_astropy_2018},
Scipy \citep{virtanen_scipy_2020},
Numpy \citep{harris_array_2020},
pandas \citep{reback_pandas-devpandas_2022},
Pyregions \citep{bradley_astropyregions_2022},
Matplotlib \citep{hunter_matplotlib_2007},
DoPHOT \citep{schechter_dophot_1993},
SWarp \citep{bertin_terapix_2002},
DECam Community Pipeline \citep{valdes_decam_2014}
}

\facility{IRSA, Spitzer}

\bibliography{references}{}

\begin{thebibliography}{}
\expandafter\ifx\csname natexlab\endcsname\relax\def\natexlab#1{#1}\fi
\providecommand{\url}[1]{\href{#1}{#1}}
\providecommand{\dodoi}[1]{doi:~\href{http://doi.org/#1}{\nolinkurl{#1}}}
\providecommand{\doeprint}[1]{\href{http://ascl.net/#1}{\nolinkurl{http://ascl.net/#1}}}
\providecommand{\doarXiv}[1]{\href{https://arxiv.org/abs/#1}{\nolinkurl{https://arxiv.org/abs/#1}}}

\bibitem[{Bertin {et~al.}(2002)Bertin, Mellier, Radovich, Missonnier, Didelon,
  \& Morin}]{bertin_terapix_2002}
Bertin, E., Mellier, Y., Radovich, M., {et~al.} 2002, 281, 228.
\newblock \url{https://ui.adsabs.harvard.edu/abs/2002ASPC..281..228B}

\bibitem[{Bianco {et~al.}(2011)Bianco, Howell, Sullivan, Conley, Kasen,
  González-Gaitán, Guy, Astier, Balland, Carlberg, Fouchez, Fourmanoit,
  Hardin, Hook, Lidman, Pain, Palanque-Delabrouille, Perlmutter, Perrett,
  Pritchet, Regnault, Rich, \& Ruhlmann-Kleider}]{bianco_constraining_2011}
Bianco, F.~B., Howell, D.~A., Sullivan, M., {et~al.} 2011, The Astrophysical
  Journal, 741, 20, \dodoi{10.1088/0004-637X/741/1/20}

\bibitem[{Bloom {et~al.}(2012)Bloom, Kasen, Shen, Nugent, Butler, Graham,
  Howell, Kolb, Holmes, Haswell, Burwitz, Rodriguez, \&
  Sullivan}]{bloom_compact_2012}
Bloom, J.~S., Kasen, D., Shen, K.~J., {et~al.} 2012, The Astrophysical Journal,
  744, L17, \dodoi{10.1088/2041-8205/744/2/L17}

\bibitem[{Bradley {et~al.}(2022)Bradley, Deil, Patra, Ginsburg, Robitaille,
  Sipőcz, King, Lim, Singer, Val-Borro, Jenness, Baumann, {Yash-10}, Donath,
  Tollerud, Lee, Leinweber, \& Vinícius}]{bradley_astropyregions_2022}
Bradley, L., Deil, C., Patra, S., {et~al.} 2022, astropy/regions: v0.5,
  Zenodo, \dodoi{10.5281/zenodo.5826359}

\bibitem[{Brown(2015)}]{brown_hypervelocity_2015}
Brown, W.~R. 2015, Annual Review of Astronomy and Astrophysics, 53, 15,
  \dodoi{10.1146/annurev-astro-082214-122230}

\bibitem[{{Capak; Peter}(2019)}]{https://doi.org/10.26131/irsa3}
{Capak; Peter}. 2019, Spitzer enhanced imaging products ({SEIP}) source list,
  IPAC, \dodoi{10.26131/IRSA3}

\bibitem[{Cartier {et~al.}(2017)Cartier, Sullivan, Firth, Pignata, Mazzali,
  Maguire, Childress, Arcavi, Ashall, Bassett, Crawford, Frohmaier, Galbany,
  Gal-Yam, Hosseinzadeh, Howell, Inserra, Johansson, Kasai, McCully, Prajs,
  Prentice, Schulze, Smartt, Smith, Smith, Valenti, \&
  Young}]{cartier_early_2017}
Cartier, R., Sullivan, M., Firth, R.~E., {et~al.} 2017, Monthly Notices of the
  Royal Astronomical Society, 464, 4476, \dodoi{10.1093/mnras/stw2678}

\bibitem[{Colgate \& McKee(1969)}]{colgate_early_1969}
Colgate, S.~A., \& McKee, C. 1969, The Astrophysical Journal, 157, 623,
  \dodoi{10.1086/150102}

\bibitem[{Collaboration {et~al.}(2016)Collaboration, Prusti, de~Bruijne, Brown,
  Vallenari, Babusiaux, Bailer-Jones, Bastian, Biermann, Evans, Eyer, Jansen,
  Jordi, Klioner, Lammers, Lindegren, Luri, Mignard, Milligan, Panem,
  Poinsignon, Pourbaix, Randich, Sarri, Sartoretti, Siddiqui, Soubiran,
  Valette, van Leeuwen, Walton, Aerts, Arenou, Cropper, Drimmel, Høg, Katz,
  Lattanzi, O'Mullane, Grebel, Holland, Huc, Passot, Bramante, Cacciari,
  Castañeda, Chaoul, Cheek, De~Angeli, Fabricius, Guerra, Hernández,
  Jean-Antoine-Piccolo, Masana, Messineo, Mowlavi, Nienartowicz,
  Ordóñez-Blanco, Panuzzo, Portell, Richards, Riello, Seabroke, Tanga,
  Thévenin, Torra, Els, Gracia-Abril, Comoretto, Garcia-Reinaldos, Lock,
  Mercier, Altmann, Andrae, Astraatmadja, Bellas-Velidis, Benson, Berthier,
  Blomme, Busso, Carry, Cellino, Clementini, Cowell, Creevey, Cuypers,
  Davidson, De~Ridder, de~Torres, Delchambre, Dell'Oro, Ducourant, Frémat,
  García-Torres, Gosset, Halbwachs, Hambly, Harrison, Hauser, Hestroffer,
  Hodgkin, Huckle, Hutton, Jasniewicz, Jordan, Kontizas, Korn, Lanzafame,
  Manteiga, Moitinho, Muinonen, Osinde, Pancino, Pauwels, Petit, Recio-Blanco,
  Robin, Sarro, Siopis, Smith, Smith, Sozzetti, Thuillot, van Reeven, Viala,
  Abbas, Abreu~Aramburu, Accart, Aguado, Allan, Allasia, Altavilla, Álvarez,
  Alves, Anderson, Andrei, Anglada~Varela, Antiche, Antoja, Antón, Arcay,
  Atzei, Ayache, Bach, Baker, Balaguer-Núñez, Barache, Barata, Barbier,
  Barblan, Baroni, Barrado~y Navascués, Barros, Barstow, Becciani, Bellazzini,
  Bellei, Bello~García, Belokurov, Bendjoya, Berihuete, Bianchi, Bienaymé,
  Billebaud, Blagorodnova, Blanco-Cuaresma, Boch, Bombrun, Borrachero,
  Bouquillon, Bourda, Bouy, Bragaglia, Breddels, Brouillet, Brüsemeister,
  Bucciarelli, Budnik, Burgess, Burgon, Burlacu, Busonero, Buzzi, Caffau,
  Cambras, Campbell, Cancelliere, Cantat-Gaudin, Carlucci, Carrasco,
  Castellani, Charlot, Charnas, Charvet, Chassat, Chiavassa, Clotet, Cocozza,
  Collins, Collins, Costigan, Crifo, Cross, Crosta, Crowley, Dafonte, Damerdji,
  Dapergolas, David, David, De~Cat, de~Felice, de~Laverny, De~Luise, De~March,
  de~Martino, de~Souza, Debosscher, del Pozo, Delbo, Delgado, Delgado,
  di~Marco, Di~Matteo, Diakite, Distefano, Dolding, Dos~Anjos, Drazinos,
  Durán, Dzigan, Ecale, Edvardsson, Enke, Erdmann, Escolar, Espina, Evans,
  Eynard~Bontemps, Fabre, Fabrizio, Faigler, Falcão, Farràs~Casas, Faye,
  Federici, Fedorets, Fernández-Hernández, Fernique, Fienga, Figueras,
  Filippi, Findeisen, Fonti, Fouesneau, Fraile, Fraser, Fuchs, Furnell, Gai,
  Galleti, Galluccio, Garabato, García-Sedano, Garé, Garofalo, Garralda,
  Gavras, Gerssen, Geyer, Gilmore, Girona, Giuffrida, Gomes, González-Marcos,
  González-Núñez, González-Vidal, Granvik, Guerrier, Guillout, Guiraud,
  Gúrpide, Gutiérrez-Sánchez, Guy, Haigron, Hatzidimitriou, Haywood, Heiter,
  Helmi, Hobbs, Hofmann, Holl, Holland, Hunt, Hypki, Icardi, Irwin, Jevardat~de
  Fombelle, Jofré, Jonker, Jorissen, Julbe, Karampelas, Kochoska, Kohley,
  Kolenberg, Kontizas, Koposov, Kordopatis, Koubsky, Kowalczyk, Krone-Martins,
  Kudryashova, Kull, Bachchan, Lacoste-Seris, Lanza, Lavigne,
  Le~Poncin-Lafitte, Lebreton, Lebzelter, Leccia, Leclerc, Lecoeur-Taibi,
  Lemaitre, Lenhardt, Leroux, Liao, Licata, Lindstrøm, Lister, Livanou, Lobel,
  Löffler, López, Lopez-Lozano, Lorenz, Loureiro, MacDonald,
  Magalhães~Fernandes, Managau, Mann, Mantelet, Marchal, Marchant, Marconi,
  Marie, Marinoni, Marrese, Marschalkó, Marshall, Martín-Fleitas, Martino,
  Mary, Matijevič, Mazeh, McMillan, Messina, Mestre, Michalik, Millar,
  Miranda, Molina, Molinaro, Molinaro, Molnár, Moniez, Montegriffo, Monteiro,
  Mor, Mora, Morbidelli, Morel, Morgenthaler, Morley, Morris, Mulone, Muraveva,
  Musella, Narbonne, Nelemans, Nicastro, Noval, Ordénovic, Ordieres-Meré,
  Osborne, Pagani, Pagano, Pailler, Palacin, Palaversa, Parsons, Paulsen,
  Pecoraro, Pedrosa, Pentikäinen, Pereira, Pichon, Piersimoni, Pineau, Plachy,
  Plum, Poujoulet, Prša, Pulone, Ragaini, Rago, Rambaux, Ramos-Lerate,
  Ranalli, Rauw, Read, Regibo, Renk, Reylé, Ribeiro, Rimoldini, Ripepi, Riva,
  Rixon, Roelens, Romero-Gómez, Rowell, Royer, Rudolph, Ruiz-Dern, Sadowski,
  Sagristà~Sellés, Sahlmann, Salgado, Salguero, Sarasso, Savietto, Schnorhk,
  Schultheis, Sciacca, Segol, Segovia, Segransan, Serpell, Shih, Smareglia,
  Smart, Smith, Solano, Solitro, Sordo, Soria~Nieto, Souchay, Spagna, Spoto,
  Stampa, Steele, Steidelmüller, Stephenson, Stoev, Suess, Süveges, Surdej,
  Szabados, Szegedi-Elek, Tapiador, Taris, Tauran, Taylor, Teixeira, Terrett,
  Tingley, Trager, Turon, Ulla, Utrilla, Valentini, van Elteren, Van~Hemelryck,
  van Leeuwen, Varadi, Vecchiato, Veljanoski, Via, Vicente, Vogt, Voss,
  Votruba, Voutsinas, Walmsley, Weiler, Weingrill, Werner, Wevers, Whitehead,
  Wyrzykowski, Yoldas, Žerjal, Zucker, Zurbach, Zwitter, Alecu, Allen,
  Allende~Prieto, Amorim, Anglada-Escudé, Arsenijevic, Azaz, Balm, Beck,
  Bernstein, Bigot, Bijaoui, Blasco, Bonfigli, Bono, Boudreault, Bressan,
  Brown, Brunet, Bunclark, Buonanno, Butkevich, Carret, Carrion, Chemin,
  Chéreau, Corcione, Darmigny, de~Boer, de~Teodoro, de~Zeeuw, Delle~Luche,
  Domingues, Dubath, Fodor, Frézouls, Fries, Fustes, Fyfe, Gallardo, Gallegos,
  Gardiol, Gebran, Gomboc, Gómez, Grux, Gueguen, Heyrovsky, Hoar, Iannicola,
  Isasi~Parache, Janotto, Joliet, Jonckheere, Keil, Kim, Klagyivik, Klar,
  Knude, Kochukhov, Kolka, Kos, Kutka, Lainey, LeBouquin, Liu, Loreggia,
  Makarov, Marseille, Martayan, Martinez-Rubi, Massart, Meynadier, Mignot,
  Munari, Nguyen, Nordlander, Ocvirk, O'Flaherty, Olias~Sanz, Ortiz, Osorio,
  Oszkiewicz, Ouzounis, Palmer, Park, Pasquato, Peltzer, Peralta, Péturaud,
  Pieniluoma, Pigozzi, Poels, Prat, Prod'homme, Raison, Rebordao, Risquez,
  Rocca-Volmerange, Rosen, Ruiz-Fuertes, Russo, Sembay, Serraller~Vizcaino,
  Short, Siebert, Silva, Sinachopoulos, Slezak, Soffel, Sosnowska, Straižys,
  ter Linden, Terrell, Theil, Tiede, Troisi, Tsalmantza, Tur, Vaccari, Vachier,
  Valles, Van~Hamme, Veltz, Virtanen, Wallut, Wichmann, Wilkinson, Ziaeepour,
  \& Zschocke}]{collaboration_gaia_2016}
Collaboration, G., Prusti, T., de~Bruijne, J. H.~J., {et~al.} 2016, Astronomy
  and Astrophysics, 595, A1, \dodoi{10.1051/0004-6361/201629272}

\bibitem[{Collaboration {et~al.}(2018{\natexlab{a}})Collaboration, Brown,
  Vallenari, Prusti, de~Bruijne, Babusiaux, Bailer-Jones, Biermann, Evans,
  Eyer, Jansen, Jordi, Klioner, Lammers, Lindegren, Luri, Mignard, Panem,
  Pourbaix, Randich, Sartoretti, Siddiqui, Soubiran, van Leeuwen, Walton,
  Arenou, Bastian, Cropper, Drimmel, Katz, Lattanzi, Bakker, Cacciari,
  Castañeda, Chaoul, Cheek, De~Angeli, Fabricius, Guerra, Holl, Masana,
  Messineo, Mowlavi, Nienartowicz, Panuzzo, Portell, Riello, Seabroke, Tanga,
  Thévenin, Gracia-Abril, Comoretto, Garcia-Reinaldos, Teyssier, Altmann,
  Andrae, Audard, Bellas-Velidis, Benson, Berthier, Blomme, Burgess, Busso,
  Carry, Cellino, Clementini, Clotet, Creevey, Davidson, De~Ridder, Delchambre,
  Dell'Oro, Ducourant, Fernández-Hernández, Fouesneau, Frémat, Galluccio,
  García-Torres, González-Núñez, González-Vidal, Gosset, Guy, Halbwachs,
  Hambly, Harrison, Hernández, Hestroffer, Hodgkin, Hutton, Jasniewicz,
  Jean-Antoine-Piccolo, Jordan, Korn, Krone-Martins, Lanzafame, Lebzelter,
  Löffler, Manteiga, Marrese, Martín-Fleitas, Moitinho, Mora, Muinonen,
  Osinde, Pancino, Pauwels, Petit, Recio-Blanco, Richards, Rimoldini, Robin,
  Sarro, Siopis, Smith, Sozzetti, Süveges, Torra, van Reeven, Abbas,
  Abreu~Aramburu, Accart, Aerts, Altavilla, Álvarez, Alvarez, Alves, Anderson,
  Andrei, Anglada~Varela, Antiche, Antoja, Arcay, Astraatmadja, Bach, Baker,
  Balaguer-Núñez, Balm, Barache, Barata, Barbato, Barblan, Barklem, Barrado,
  Barros, Barstow, Bartholomé~Muñoz, Bassilana, Becciani, Bellazzini,
  Berihuete, Bertone, Bianchi, Bienaymé, Blanco-Cuaresma, Boch, Boeche,
  Bombrun, Borrachero, Bossini, Bouquillon, Bourda, Bragaglia, Bramante,
  Breddels, Bressan, Brouillet, Brüsemeister, Brugaletta, Bucciarelli,
  Burlacu, Busonero, Butkevich, Buzzi, Caffau, Cancelliere, Cannizzaro,
  Cantat-Gaudin, Carballo, Carlucci, Carrasco, Casamiquela, Castellani,
  Castro-Ginard, Charlot, Chemin, Chiavassa, Cocozza, Costigan, Cowell, Crifo,
  Crosta, Crowley, Cuypers, Dafonte, Damerdji, Dapergolas, David, David,
  de~Laverny, De~Luise, De~March, de~Martino, de~Souza, de~Torres, Debosscher,
  del Pozo, Delbo, Delgado, Delgado, Di~Matteo, Diakite, Diener, Distefano,
  Dolding, Drazinos, Durán, Edvardsson, Enke, Eriksson, Esquej,
  Eynard~Bontemps, Fabre, Fabrizio, Faigler, Falcão, Farràs~Casas, Federici,
  Fedorets, Fernique, Figueras, Filippi, Findeisen, Fonti, Fraile, Fraser,
  Frézouls, Gai, Galleti, Garabato, García-Sedano, Garofalo, Garralda, Gavel,
  Gavras, Gerssen, Geyer, Giacobbe, Gilmore, Girona, Giuffrida, Glass, Gomes,
  Granvik, Gueguen, Guerrier, Guiraud, Gutiérrez-Sánchez, Haigron,
  Hatzidimitriou, Hauser, Haywood, Heiter, Helmi, Heu, Hilger, Hobbs, Hofmann,
  Holland, Huckle, Hypki, Icardi, Janßen, Jevardat~de Fombelle, Jonker,
  Juhász, Julbe, Karampelas, Kewley, Klar, Kochoska, Kohley, Kolenberg,
  Kontizas, Kontizas, Koposov, Kordopatis, Kostrzewa-Rutkowska, Koubsky,
  Lambert, Lanza, Lasne, Lavigne, Le~Fustec, Le~Poncin-Lafitte, Lebreton,
  Leccia, Leclerc, Lecoeur-Taibi, Lenhardt, Leroux, Liao, Licata, Lindstrøm,
  Lister, Livanou, Lobel, López, Managau, Mann, Mantelet, Marchal, Marchant,
  Marconi, Marinoni, Marschalkó, Marshall, Martino, Marton, Mary, Massari,
  Matijevič, Mazeh, McMillan, Messina, Michalik, Millar, Molina, Molinaro,
  Molnár, Montegriffo, Mor, Morbidelli, Morel, Morris, Mulone, Muraveva,
  Musella, Nelemans, Nicastro, Noval, O'Mullane, Ordénovic, Ordóñez-Blanco,
  Osborne, Pagani, Pagano, Pailler, Palacin, Palaversa, Panahi, Pawlak,
  Piersimoni, Pineau, Plachy, Plum, Poggio, Poujoulet, Prša, Pulone, Racero,
  Ragaini, Rambaux, Ramos-Lerate, Regibo, Reylé, Riclet, Ripepi, Riva, Rivard,
  Rixon, Roegiers, Roelens, Romero-Gómez, Rowell, Royer, Ruiz-Dern, Sadowski,
  Sagristà~Sellés, Sahlmann, Salgado, Salguero, Sanna, Santana-Ros, Sarasso,
  Savietto, Schultheis, Sciacca, Segol, Segovia, Ségransan, Shih, Siltala,
  Silva, Smart, Smith, Solano, Solitro, Sordo, Soria~Nieto, Souchay, Spagna,
  Spoto, Stampa, Steele, Steidelmüller, Stephenson, Stoev, Suess, Surdej,
  Szabados, Szegedi-Elek, Tapiador, Taris, Tauran, Taylor, Teixeira, Terrett,
  Teyssandier, Thuillot, Titarenko, Torra~Clotet, Turon, Ulla, Utrilla, Uzzi,
  Vaillant, Valentini, Valette, van Elteren, Van~Hemelryck, van Leeuwen,
  Vaschetto, Vecchiato, Veljanoski, Viala, Vicente, Vogt, von Essen, Voss,
  Votruba, Voutsinas, Walmsley, Weiler, Wertz, Wevers, Wyrzykowski, Yoldas,
  Žerjal, Ziaeepour, Zorec, Zschocke, Zucker, Zurbach, \&
  Zwitter}]{collaboration_gaia_2018}
Collaboration, G., Brown, A. G.~A., Vallenari, A., {et~al.} 2018{\natexlab{a}},
  Astronomy and Astrophysics, 616, A1, \dodoi{10.1051/0004-6361/201833051}

\bibitem[{Collaboration {et~al.}(2018{\natexlab{b}})Collaboration,
  Price-Whelan, Sipőcz, Günther, Lim, Crawford, Conseil, Shupe, Craig,
  Dencheva, Ginsburg, VanderPlas, Bradley, Pérez-Suárez, de~Val-Borro,
  Aldcroft, Cruz, Robitaille, Tollerud, Ardelean, Babej, Bachetti, Bakanov,
  Bamford, Barentsen, Barmby, Baumbach, Berry, Biscani, Boquien, Bostroem,
  Bouma, Brammer, Bray, Breytenbach, Buddelmeijer, Burke, Calderone,
  Rodríguez, Cara, Cardoso, Cheedella, Copin, Crichton, DÁvella, Deil,
  Depagne, Dietrich, Donath, Droettboom, Earl, Erben, Fabbro, Ferreira,
  Finethy, Fox, Garrison, Gibbons, Goldstein, Gommers, Greco, Greenfield,
  Groener, Grollier, Hagen, Hirst, Homeier, Horton, Hosseinzadeh, Hu, Hunkeler,
  Ivezić, Jain, Jenness, Kanarek, Kendrew, Kern, Kerzendorf, Khvalko, King,
  Kirkby, Kulkarni, Kumar, Lee, Lenz, Littlefair, Ma, Macleod, Mastropietro,
  McCully, Montagnac, Morris, Mueller, Mumford, Muna, Murphy, Nelson, Nguyen,
  Ninan, Nöthe, Ogaz, Oh, Parejko, Parley, Pascual, Patil, Patil, Plunkett,
  Prochaska, Rastogi, Janga, Sabater, Sakurikar, Seifert, Sherbert,
  Sherwood-Taylor, Shih, Sick, Silbiger, Singanamalla, Singer, Sladen, Sooley,
  Sornarajah, Streicher, Teuben, Thomas, Tremblay, Turner, Terrón, van
  Kerkwijk, de~la Vega, Watkins, Weaver, Whitmore, Woillez, \&
  Zabalza}]{collaboration_astropy_2018}
Collaboration, T.~A., Price-Whelan, A.~M., Sipőcz, B.~M., {et~al.}
  2018{\natexlab{b}}, \dodoi{10.3847/1538-3881/aabc4f}

\bibitem[{Dan {et~al.}(2011)Dan, Rosswog, Guillochon, \&
  Ramirez-Ruiz}]{dan_prelude_2011}
Dan, M., Rosswog, S., Guillochon, J., \& Ramirez-Ruiz, E. 2011, The
  Astrophysical Journal, 737, 89, \dodoi{10.1088/0004-637X/737/2/89}

\bibitem[{Diehl(2012)}]{diehl_dark_2012}
Diehl, T. 2012, Physics Procedia, 37, 1332, \dodoi{10.1016/j.phpro.2012.02.472}

\bibitem[{Fausnaugh {et~al.}(2021)Fausnaugh, Vallely, Kochanek, Shappee,
  Stanek, Tucker, Ricker, Vanderspek, Latham, Seager, Winn, Jenkins,
  Berta-Thompson, Daylan, Doty, Fűrész, Levine, Morris, Pál, Sha, Ting, \&
  Wohler}]{fausnaugh_early-time_2021}
Fausnaugh, M.~M., Vallely, P.~J., Kochanek, C.~S., {et~al.} 2021, The
  Astrophysical Journal, 908, 51, \dodoi{10.3847/1538-4357/abcd42}

\bibitem[{Fazio {et~al.}(2004)Fazio, Hora, Allen, Ashby, Barmby, Deutsch,
  Huang, Kleiner, Marengo, Megeath, Melnick, Pahre, Patten, Polizotti, Smith,
  Taylor, Wang, Willner, Hoffmann, Pipher, Forrest, McMurty, McCreight,
  McKelvey, McMurray, Koch, Moseley, Arendt, Mentzell, Marx, Losch, Mayman,
  Eichhorn, Krebs, Jhabvala, Gezari, Fixsen, Flores, Shakoorzadeh, Jungo,
  Hakun, Workman, Karpati, Kichak, Whitley, Mann, Tollestrup, Eisenhardt,
  Stern, Gorjian, Bhattacharya, Carey, Nelson, Glaccum, Lacy, Lowrance, Laine,
  Reach, Stauffer, Surace, Wilson, Wright, Hoffman, Domingo, \&
  Cohen}]{fazio_infrared_2004}
Fazio, G.~G., Hora, J.~L., Allen, L.~E., {et~al.} 2004, The Astrophysical
  Journal Supplement Series, 154, 10, \dodoi{10.1086/422843}

\bibitem[{Flaugher {et~al.}(2015)Flaugher, Diehl, Honscheid, Abbott, Alvarez,
  Angstadt, Annis, Antonik, Ballester, Beaufore, Bernstein, Bernstein, Bigelow,
  Bonati, Boprie, Brooks, Buckley-Geer, Campa, Cardiel-Sas, Castander,
  Castilla, Cease, Cela-Ruiz, Chappa, Chi, Cooper, da~Costa, Dede, Derylo,
  DePoy, de~Vicente, Doel, Drlica-Wagner, Eiting, Elliott, Emes, Estrada, Neto,
  Finley, Flores, Frieman, Gerdes, Gladders, Gregory, Gutierrez, Hao, Holland,
  Holm, Huffman, Jackson, James, Jonas, Karcher, Karliner, Kent, Kessler,
  Kozlovsky, Kron, Kubik, Kuehn, Kuhlmann, Kuk, Lahav, Lathrop, Lee, Levi,
  Lewis, Li, Mandrichenko, Marshall, Martinez, Merritt, Miquel, Munoz, Neilsen,
  Nichol, Nord, Ogando, Olsen, Palio, Patton, Peoples, Plazas, Rauch, Reil,
  Rheault, Roe, Rogers, Roodman, Sanchez, Scarpine, Schindler, Schmidt,
  Schmitt, Schubnell, Schultz, Schurter, Scott, Serrano, Shaw, Smith,
  Soares-Santos, Stefanik, Stuermer, Suchyta, Sypniewski, Tarle, Thaler, Tighe,
  Tran, Tucker, Walker, Wang, Watson, Weaverdyck, Wester, Woods, \&
  Yanny}]{flaugher_dark_2015}
Flaugher, B., Diehl, H.~T., Honscheid, K., {et~al.} 2015, The Astronomical
  Journal, 150, 150, \dodoi{10.1088/0004-6256/150/5/150}

\bibitem[{{Gaia Collaboration} {et~al.}(2021){Gaia Collaboration}, Brown,
  Vallenari, Prusti, de~Bruijne, Babusiaux, Biermann, Creevey, Evans, Eyer,
  Hutton, Jansen, Jordi, Klioner, Lammers, Lindegren, Luri, Mignard, Panem,
  Pourbaix, Randich, Sartoretti, Soubiran, Walton, Arenou, Bailer-Jones,
  Bastian, Cropper, Drimmel, Katz, Lattanzi, van Leeuwen, Bakker, Cacciari,
  Castañeda, De~Angeli, Ducourant, Fabricius, Fouesneau, Frémat, Guerra,
  Guerrier, Guiraud, Jean-Antoine~Piccolo, Masana, Messineo, Mowlavi, Nicolas,
  Nienartowicz, Pailler, Panuzzo, Riclet, Roux, Seabroke, Sordo, Tanga,
  Thévenin, Gracia-Abril, Portell, Teyssier, Altmann, Andrae, Bellas-Velidis,
  Benson, Berthier, Blomme, Brugaletta, Burgess, Busso, Carry, Cellino, Cheek,
  Clementini, Damerdji, Davidson, Delchambre, Dell’Oro,
  Fernández-Hernández, Galluccio, García-Lario, Garcia-Reinaldos,
  González-Núñez, Gosset, Haigron, Halbwachs, Hambly, Harrison,
  Hatzidimitriou, Heiter, Hernández, Hestroffer, Hodgkin, Holl, Janßen,
  Jevardat~de Fombelle, Jordan, Krone-Martins, Lanzafame, Löffler, Lorca,
  Manteiga, Marchal, Marrese, Moitinho, Mora, Muinonen, Osborne, Pancino,
  Pauwels, Petit, Recio-Blanco, Richards, Riello, Rimoldini, Robin, Roegiers,
  Rybizki, Sarro, Siopis, Smith, Sozzetti, Ulla, Utrilla, van Leeuwen, van
  Reeven, Abbas, Abreu~Aramburu, Accart, Aerts, Aguado, Ajaj, Altavilla,
  Álvarez, Álvarez Cid-Fuentes, Alves, Anderson, Anglada~Varela, Antoja,
  Audard, Baines, Baker, Balaguer-Núñez, Balbinot, Balog, Barache, Barbato,
  Barros, Barstow, Bartolomé, Bassilana, Bauchet, Baudesson-Stella, Becciani,
  Bellazzini, Bernet, Bertone, Bianchi, Blanco-Cuaresma, Boch, Bombrun,
  Bossini, Bouquillon, Bragaglia, Bramante, Breedt, Bressan, Brouillet,
  Bucciarelli, Burlacu, Busonero, Butkevich, Buzzi, Caffau, Cancelliere,
  Cánovas, Cantat-Gaudin, Carballo, Carlucci, Carnerero, Carrasco,
  Casamiquela, Castellani, Castro-Ginard, Castro~Sampol, Chaoul, Charlot,
  Chemin, Chiavassa, Cioni, Comoretto, Cooper, Cornez, Cowell, Crifo, Crosta,
  Crowley, Dafonte, Dapergolas, David, David, de~Laverny, De~Luise, De~March,
  De~Ridder, de~Souza, de~Teodoro, de~Torres, del Peloso, del Pozo, Delbo,
  Delgado, Delgado, Delisle, Di~Matteo, Diakite, Diener, Distefano, Dolding,
  Eappachen, Edvardsson, Enke, Esquej, Fabre, Fabrizio, Faigler, Fedorets,
  Fernique, Fienga, Figueras, Fouron, Fragkoudi, Fraile, Franke, Gai, Garabato,
  Garcia-Gutierrez, García-Torres, Garofalo, Gavras, Gerlach, Geyer, Giacobbe,
  Gilmore, Girona, Giuffrida, Gomel, Gomez, Gonzalez-Santamaria,
  González-Vidal, Granvik, Gutiérrez-Sánchez, Guy, Hauser, Haywood, Helmi,
  Hidalgo, Hilger, Hładczuk, Hobbs, Holland, Huckle, Jasniewicz, Jonker,
  Juaristi~Campillo, Julbe, Karbevska, Kervella, Khanna, Kochoska, Kontizas,
  Kordopatis, Korn, Kostrzewa-Rutkowska, Kruszyńska, Lambert, Lanza, Lasne,
  Le~Campion, Le~Fustec, Lebreton, Lebzelter, Leccia, Leclerc, Lecoeur-Taibi,
  Liao, Licata, Lindstrøm, Lister, Livanou, Lobel, Madrero~Pardo, Managau,
  Mann, Marchant, Marconi, Marcos~Santos, Marinoni, Marocco, Marshall,
  Martin~Polo, Martín-Fleitas, Masip, Massari, Mastrobuono-Battisti, Mazeh,
  McMillan, Messina, Michalik, Millar, Mints, Molina, Molinaro, Molnár,
  Montegriffo, Mor, Morbidelli, Morel, Morris, Mulone, Munoz, Muraveva, Murphy,
  Musella, Noval, Ordénovic, Orrù, Osinde, Pagani, Pagano, Palaversa,
  Palicio, Panahi, Pawlak, Peñalosa~Esteller, Penttilä, Piersimoni, Pineau,
  Plachy, Plum, Poggio, Poretti, Poujoulet, Prša, Pulone, Racero, Ragaini,
  Rainer, Raiteri, Rambaux, Ramos, Ramos-Lerate, Re~Fiorentin, Regibo, Reylé,
  Ripepi, Riva, Rixon, Robichon, Robin, Roelens, Rohrbasser, Romero-Gómez,
  Rowell, Royer, Rybicki, Sadowski, Sagristà~Sellés, Sahlmann, Salgado,
  Salguero, Samaras, Sanchez~Gimenez, Sanna, Santoveña, Sarasso, Schultheis,
  Sciacca, Segol, Segovia, Ségransan, Semeux, Shahaf, Siddiqui, Siebert,
  Siltala, Slezak, Smart, Solano, Solitro, Souami, Souchay, Spagna, Spoto,
  Steele, Steidelmüller, Stephenson, Süveges, Szabados, Szegedi-Elek, Taris,
  Tauran, Taylor, Teixeira, Thuillot, Tonello, Torra, Torra, Turon, Unger,
  Vaillant, van Dillen, Vanel, Vecchiato, Viala, Vicente, Voutsinas, Weiler,
  Wevers, Wyrzykowski, Yoldas, Yvard, Zhao, Zorec, Zucker, Zurbach, \&
  Zwitter}]{gaia_collaboration_gaia_2021}
{Gaia Collaboration}, Brown, A. G.~A., Vallenari, A., {et~al.} 2021, Astronomy
  \& Astrophysics, 649, A1, \dodoi{10.1051/0004-6361/202039657}

\bibitem[{Generozov \& Perets(2022)}]{generozov_constraints_2022}
Generozov, A., \& Perets, H.~B. 2022, Monthly Notices of the Royal Astronomical
  Society, \dodoi{10.1093/mnras/stac1108}

\bibitem[{Graham {et~al.}(2017)Graham, Kumar, Hosseinzadeh, Hiramatsu, Arcavi,
  Howell, Valenti, Sand, Parrent, McCully, \&
  Filippenko}]{graham_nebular-phase_2017}
Graham, M.~L., Kumar, S., Hosseinzadeh, G., {et~al.} 2017, Monthly Notices of
  the Royal Astronomical Society, 472, 3437, \dodoi{10.1093/mnras/stx2224}

\bibitem[{Green(2019)}]{green_revised_2019}
Green, D.~A. 2019, Journal of Astrophysics and Astronomy, 40, 36,
  \dodoi{10.1007/s12036-019-9601-6}

\bibitem[{Guillochon {et~al.}(2010)Guillochon, Dan, Ramirez-Ruiz, \&
  Rosswog}]{guillochon_surface_2010}
Guillochon, J., Dan, M., Ramirez-Ruiz, E., \& Rosswog, S. 2010, The
  Astrophysical Journal, 709, L64, \dodoi{10.1088/2041-8205/709/1/L64}

\bibitem[{Guo {et~al.}(2021)Guo, Chen, Yuan, Huang, Liu, Yang, Li, Sun, \&
  Liu}]{guo_three-dimensional_2021}
Guo, H.~L., Chen, B.~Q., Yuan, H.~B., {et~al.} 2021, The Astrophysical Journal,
  906, 47, \dodoi{10.3847/1538-4357/abc68a}

\bibitem[{Harris {et~al.}(2020)Harris, Millman, van~der Walt, Gommers,
  Virtanen, Cournapeau, Wieser, Taylor, Berg, Smith, Kern, Picus, Hoyer, van
  Kerkwijk, Brett, Haldane, del Río, Wiebe, Peterson, Gérard-Marchant,
  Sheppard, Reddy, Weckesser, Abbasi, Gohlke, \& Oliphant}]{harris_array_2020}
Harris, C.~R., Millman, K.~J., van~der Walt, S.~J., {et~al.} 2020, Nature, 585,
  357, \dodoi{10.1038/s41586-020-2649-2}

\bibitem[{Hayden {et~al.}(2010)Hayden, Garnavich, Kessler, Frieman, Jha,
  Cinabro, Dilday, Kasen, Marriner, Nichol, Riess, Sako, Schneider, Smith,
  Sollerman, \& Bassett}]{hayden_rise_2010}
Hayden, B.~T., Garnavich, P.~M., Kessler, R., {et~al.} 2010, The Astrophysical
  Journal, 712, 350, \dodoi{10.1088/0004-637X/712/1/350}

\bibitem[{Hernandez {et~al.}(2012)Hernandez, Ruiz-Lapuente, Tabernero, Montes,
  Canal, Mendez, \& Bedin}]{hernandez_no_2012}
Hernandez, J. I.~G., Ruiz-Lapuente, P., Tabernero, H.~M., {et~al.} 2012,
  \dodoi{10.1038/nature11447}

\bibitem[{Hernández {et~al.}(2009)Hernández, Ruiz-Lapuente, Filippenko,
  Foley, Gal-Yam, \& Simon}]{hernandez_chemical_2009}
Hernández, J. I.~G., Ruiz-Lapuente, P., Filippenko, A.~V., {et~al.} 2009, The
  Astrophysical Journal, 691, 1, \dodoi{10.1088/0004-637X/691/1/1}

\bibitem[{Hillebrandt \& Niemeyer(2000)}]{hillebrandt_type_2000}
Hillebrandt, W., \& Niemeyer, J.~C. 2000, Annual Review of Astronomy and
  Astrophysics, 38, 191, \dodoi{10.1146/annurev.astro.38.1.191}

\bibitem[{Hills(1988)}]{hills_hyper-velocity_1988}
Hills, J.~G. 1988, Nature, 331, 687, \dodoi{10.1038/331687a0}

\bibitem[{Holmbo {et~al.}(2019)Holmbo, Stritzinger, Shappee, Tucker, Zheng,
  Ashall, Phillips, Contreras, Filippenko, Hoeflich, Huber, Wang, Zhang, Anais,
  Baron, Burns, Campillay, Castellon, Corco, Hsiao, Krisciunas, Morrell,
  Nielsen, Persson, Piro, Taddia, Tomasella, Zhang, \&
  Zhao}]{holmbo_first_2019}
Holmbo, S., Stritzinger, M.~D., Shappee, B.~J., {et~al.} 2019, Astronomy \&
  Astrophysics, 627, A174, \dodoi{10.1051/0004-6361/201834389}

\bibitem[{Hunter(2007)}]{hunter_matplotlib_2007}
Hunter, J.~D. 2007, Computing in Science \& Engineering, 9, 90,
  \dodoi{10.1109/MCSE.2007.55}

\bibitem[{Iben {et~al.}(1987)Iben, Nomoto, Tornambe, \&
  Tutukov}]{iben_interacting_1987}
Iben, Jr., I., Nomoto, K., Tornambe, A., \& Tutukov, A.~V. 1987, The
  Astrophysical Journal, 317, 717, \dodoi{10.1086/165318}

\bibitem[{Iben \& Tutukov(1984)}]{iben_supernovae_1984}
Iben, Jr., I., \& Tutukov, A.~V. 1984, The Astrophysical Journal Supplement
  Series, 54, 335, \dodoi{10.1086/190932}

\bibitem[{Ihara {et~al.}(2007)Ihara, Ozaki, Doi, Shigeyama, Kashikawa,
  Komiyama, \& Hattori}]{ihara_searching_2007}
Ihara, Y., Ozaki, J., Doi, M., {et~al.} 2007, Publications of the Astronomical
  Society of Japan, 59, 811, \dodoi{10.1093/pasj/59.4.811}

\bibitem[{Kerzendorf {et~al.}(2014)Kerzendorf, Childress, Scharwächter, Do, \&
  Schmidt}]{kerzendorf_reconnaissance_2014}
Kerzendorf, W.~E., Childress, M., Scharwächter, J., Do, T., \& Schmidt, B.~P.
  2014, {\textbackslash}apj, 782, 27, \dodoi{10.1088/0004-637X/782/1/27}

\bibitem[{Kerzendorf {et~al.}(2018)Kerzendorf, Long, Winkler, \&
  Do}]{kerzendorf_tycho-b_2018}
Kerzendorf, W.~E., Long, K.~S., Winkler, P.~F., \& Do, T. 2018,
  {\textbackslash}mnras, 479, 5696, \dodoi{10.1093/mnras/sty1863}

\bibitem[{Kerzendorf {et~al.}(2009)Kerzendorf, Schmidt, Asplund, Nomoto,
  Podsiadlowski, Frebel, Fesen, \& Yong}]{kerzendorf_subaru_2009}
Kerzendorf, W.~E., Schmidt, B.~P., Asplund, M., {et~al.} 2009,
  \dodoi{10.1088/0004-637X/701/2/1665}

\bibitem[{Kerzendorf {et~al.}(2012)Kerzendorf, Schmidt, Laird, Podsiadlowski,
  \& Bessell}]{kerzendorf_hunting_2012}
Kerzendorf, W.~E., Schmidt, B.~P., Laird, J.~B., Podsiadlowski, P., \& Bessell,
  M.~S. 2012, The Astrophysical Journal, 759, 7,
  \dodoi{10.1088/0004-637X/759/1/7}

\bibitem[{Kerzendorf {et~al.}(2017)Kerzendorf, Strampelli, Shen, Schwab,
  Pakmor, Do, Buchner, \& Rest}]{kerzendorf_search_2017}
Kerzendorf, W.~E., Strampelli, G., Shen, K.~J., {et~al.} 2017,
  \dodoi{10.1093/mnras/sty1357}

\bibitem[{Kobayashi {et~al.}(2020)Kobayashi, Leung, \&
  Nomoto}]{kobayashi_new_2020}
Kobayashi, C., Leung, S.-C., \& Nomoto, K. 2020, The Astrophysical Journal,
  895, 138, \dodoi{10.3847/1538-4357/ab8e44}

\bibitem[{Leonard(2007)}]{leonard_constraining_2007}
Leonard, D.~C. 2007, The Astrophysical Journal, 670, 1275,
  \dodoi{10.1086/522367}

\bibitem[{Liu {et~al.}(2021)Liu, Roepke, Zeng, \&
  Heger}]{liu_observational_2021}
Liu, Z.-W., Roepke, F.~K., Zeng, Y., \& Heger, A. 2021, Astronomy \&
  Astrophysics, 654, A103, \dodoi{10.1051/0004-6361/202141518}

\bibitem[{Livio(2000)}]{livio_progenitors_2000}
Livio, M. 2000, The {Progenitors} of {Type} {Ia} {Supernovae} (eprint:
  arXiv:astro-ph/9903264).
\newblock \url{https://ui.adsabs.harvard.edu/abs/2000tias.conf...33L}

\bibitem[{Lundqvist {et~al.}(2013)Lundqvist, Mattila, Sollerman, Kozma, Baron,
  Cox, Fransson, Leibundgut, \& Spyromilio}]{lundqvist_hydrogen_2013}
Lundqvist, P., Mattila, S., Sollerman, J., {et~al.} 2013, Monthly Notices of
  the Royal Astronomical Society, 435, 329, \dodoi{10.1093/mnras/stt1303}

\bibitem[{Lundqvist {et~al.}(2015)Lundqvist, Nyholm, Taddia, Sollerman,
  Johansson, Kozma, Lundqvist, Fransson, Garnavich, Kromer, Shappee, \&
  Goobar}]{lundqvist_no_2015}
Lundqvist, P., Nyholm, A., Taddia, F., {et~al.} 2015, Astronomy \&amp;
  Astrophysics, Volume 577, id.A39,
  {\textless}NUMPAGES{\textgreater}12{\textless}/NUMPAGES{\textgreater} pp.,
  577, A39, \dodoi{10.1051/0004-6361/201525719}

\bibitem[{Maguire {et~al.}(2016)Maguire, Taubenberger, Sullivan, \&
  Mazzali}]{maguire_searching_2016}
Maguire, K., Taubenberger, S., Sullivan, M., \& Mazzali, P.~A. 2016, Monthly
  Notices of the Royal Astronomical Society, 457, 3254,
  \dodoi{10.1093/mnras/stv2991}

\bibitem[{Maoz {et~al.}(2014)Maoz, Mannucci, \&
  Nelemans}]{maoz_observational_2014}
Maoz, D., Mannucci, F., \& Nelemans, G. 2014, Annual Review of Astronomy and
  Astrophysics, 52, 107, \dodoi{10.1146/annurev-astro-082812-141031}

\bibitem[{Marietta {et~al.}(2000)Marietta, Burrows, \&
  Fryxell}]{marietta_type_2000}
Marietta, E., Burrows, A., \& Fryxell, B. 2000, {\textbackslash}apjs, 128, 615,
  \dodoi{10.1086/313392}

\bibitem[{Marion {et~al.}(2016)Marion, Brown, Vinkó, Silverman, Sand, Challis,
  Kirshner, Wheeler, Berlind, Brown, Calkins, Camacho, Dhungana, Foley,
  Friedman, Graham, Howell, Hsiao, Irwin, Jha, Kehoe, Macri, Maeda, Mandel,
  McCully, Pandya, Rines, Wilhelmy, \& Zheng}]{marion_sn_2016}
Marion, G.~H., Brown, P.~J., Vinkó, J., {et~al.} 2016, The Astrophysical
  Journal, 820, 92, \dodoi{10.3847/0004-637X/820/2/92}

\bibitem[{Mattila {et~al.}(2005)Mattila, Lundqvist, Sollerman, Kozma, Baron,
  Fransson, Leibundgut, \& Nomoto}]{mattila_early_2005}
Mattila, S., Lundqvist, P., Sollerman, J., {et~al.} 2005, Astronomy and
  Astrophysics, 443, 649, \dodoi{10.1051/0004-6361:20052731}

\bibitem[{Miller {et~al.}(2018)Miller, Cao, Piro, Blagorodnova, Bue, Cenko,
  Dhawan, Ferretti, Fox, Fremling, Goobar, Howell, Hosseinzadeh, Kasliwal,
  Laher, Lunnan, Masci, McCully, Nugent, Sollerman, Taddia, \&
  Kulkarni}]{miller_early_2018}
Miller, A.~A., Cao, Y., Piro, A.~L., {et~al.} 2018, The Astrophysical Journal,
  852, 100, \dodoi{10.3847/1538-4357/aaa01f}

\bibitem[{Nomoto(1982)}]{nomoto_accreting_1982-1}
Nomoto, K. 1982, {\textbackslash}apj, 257, 780, \dodoi{10.1086/160031}

\bibitem[{Nomoto {et~al.}(2013)Nomoto, Kobayashi, \&
  Tominaga}]{nomoto_nucleosynthesis_2013}
Nomoto, K., Kobayashi, C., \& Tominaga, N. 2013, Annual Review of Astronomy and
  Astrophysics, 51, 457, \dodoi{10.1146/annurev-astro-082812-140956}

\bibitem[{Olling {et~al.}(2015)Olling, Mushotzky, Shaya, Rest, Garnavich,
  Tucker, Kasen, Margheim, \& Filippenko}]{olling_no_2015}
Olling, R.~P., Mushotzky, R., Shaya, E.~J., {et~al.} 2015, Nature, 521, 332,
  \dodoi{10.1038/nature14455}

\bibitem[{Pakmor {et~al.}(2010)Pakmor, Kromer, Röpke, Sim, Ruiter, \&
  Hillebrandt}]{pakmor_sub-luminous_2010}
Pakmor, R., Kromer, M., Röpke, F.~K., {et~al.} 2010, Nature, 463, 61,
  \dodoi{10.1038/nature08642}

\bibitem[{Pakmor {et~al.}(2013)Pakmor, Kromer, Taubenberger, \&
  Springel}]{pakmor_helium-ignited_2013}
Pakmor, R., Kromer, M., Taubenberger, S., \& Springel, V. 2013, The
  Astrophysical Journal, 770, L8, \dodoi{10.1088/2041-8205/770/1/L8}

\bibitem[{Pakmor {et~al.}(2008)Pakmor, Roepke, Weiss, \&
  Hillebrandt}]{pakmor_impact_2008}
Pakmor, R., Roepke, F.~K., Weiss, A., \& Hillebrandt, W. 2008, Astronomy \&
  Astrophysics, 489, 943, \dodoi{10.1051/0004-6361:200810456}

\bibitem[{Pakmor {et~al.}(2022)Pakmor, Callan, Collins, de~Mink, Holas,
  Kerzendorf, Kromer, O'Brien, Roepke, Ruiter, Seitenzahl, Shingles, Sim, \&
  Taubenberger}]{pakmor_fate_2022}
Pakmor, R., Callan, F.~P., Collins, C.~E., {et~al.} 2022, On the fate of the
  secondary white dwarf in double-degenerate double-detonation {Type} {Ia}
  supernovae, Tech. rep.
\newblock \url{https://ui.adsabs.harvard.edu/abs/2022arXiv220314990P}

\bibitem[{Pan {et~al.}(2012)Pan, Ricker, \& Taam}]{pan_impact_2012}
Pan, K.-C., Ricker, P.~M., \& Taam, R.~E. 2012, {\textbackslash}apj, 750, 151,
  \dodoi{10.1088/0004-637X/750/2/151}

\bibitem[{Pan {et~al.}(2013)Pan, Ricker, \& Taam}]{pan_evolution_2013}
---. 2013, {\textbackslash}apj, 773, 49, \dodoi{10.1088/0004-637X/773/1/49}

\bibitem[{Pankey(1962)}]{pankey_possible_1962}
Pankey, Jr., T. 1962, PhD thesis.
\newblock \url{https://ui.adsabs.harvard.edu/abs/1962PhDT........25P}

\bibitem[{Perlmutter {et~al.}(1999)Perlmutter, Aldering, Goldhaber, Knop,
  Nugent, Castro, Deustua, Fabbro, Goobar, Groom, Hook, Kim, Kim, Lee, Nunes,
  Pain, Pennypacker, Quimby, Lidman, Ellis, Irwin, McMahon, Ruiz-Lapuente,
  Walton, Schaefer, Boyle, Filippenko, Matheson, Fruchter, Panagia, Newberg,
  Couch, \& Project}]{perlmutter_measurements_1999}
Perlmutter, S., Aldering, G., Goldhaber, G., {et~al.} 1999, The Astrophysical
  Journal, 517, 565, \dodoi{10.1086/307221}

\bibitem[{Reback {et~al.}(2022)Reback, {jbrockmendel}, McKinney, Bossche,
  Augspurger, Cloud, Hawkins, Roeschke, {gfyoung}, {Sinhrks}, Klein, Hoefler,
  Petersen, Tratner, She, Ayd, Naveh, Darbyshire, Garcia, Shadrach, Schendel,
  Hayden, Saxton, Gorelli, Li, Zeitlin, Jancauskas, McMaster, Battiston, \&
  Seabold}]{reback_pandas-devpandas_2022}
Reback, J., {jbrockmendel}, McKinney, W., {et~al.} 2022, pandas-dev/pandas:
  {Pandas} 1.4.1,  Zenodo, \dodoi{10.5281/zenodo.6053272}

\bibitem[{Rest {et~al.}(2005)Rest, Stubbs, Becker, Miknaitis, Miceli,
  Covarrubias, Hawley, Smith, Suntzeff, Olsen, Prieto, Hiriart, Welch, Cook,
  Nikolaev, Huber, Prochtor, Clocchiatti, Minniti, Garg, Challis, Keller, \&
  Schmidt}]{rest_testing_2005}
Rest, A., Stubbs, C., Becker, A.~C., {et~al.} 2005, The Astrophysical Journal,
  634, 1103, \dodoi{10.1086/497060}

\bibitem[{Rest {et~al.}(2013)Rest, Scolnic, Foley, Huber, Chornock, Narayan,
  Tonry, Berger, Soderberg, Stubbs, Riess, Kirshner, Smartt, Schlafly, Rodney,
  Botticella, Brout, Challis, Czekala, Drout, Hudson, Kotak, Leibler, Lunnan,
  Marion, McCrum, Milisavljevic, Pastorello, Sanders, Smith, Stafford, Thilker,
  Valenti, Wood-Vasey, Zheng, Burgett, Chambers, Denneau, Draper, Flewelling,
  Hodapp, Kaiser, Kudritzki, Magnier, Metcalfe, Price, Sweeney, Wainscoat, \&
  Waters}]{rest_cosmological_2013}
Rest, A., Scolnic, D., Foley, R.~J., {et~al.} 2013,
  \dodoi{10.1088/0004-637X/795/1/44}

\bibitem[{Riess {et~al.}(1998)Riess, Filippenko, Challis, Clocchiatti, Diercks,
  Garnavich, Gilliland, Hogan, Jha, Kirshner, Leibundgut, Phillips, Reiss,
  Schmidt, Schommer, Smith, Spyromilio, Stubbs, Suntzeff, \&
  Tonry}]{riess_observational_1998}
Riess, A.~G., Filippenko, A.~V., Challis, P., {et~al.} 1998, The Astronomical
  Journal, 116, 1009, \dodoi{10.1086/300499}

\bibitem[{Ruiz-Lapuente(2019)}]{ruiz-lapuente_surviving_2019}
Ruiz-Lapuente, P. 2019, New Astronomy Reviews, 85, 101523,
  \dodoi{10.1016/j.newar.2019.101523}

\bibitem[{Ruiz-Lapuente {et~al.}(2018)Ruiz-Lapuente, Damiani, Bedin,
  González~Hernández, Galbany, Pritchard, Canal, \&
  Méndez}]{ruiz-lapuente_no_2018}
Ruiz-Lapuente, P., Damiani, F., Bedin, L., {et~al.} 2018, The Astrophysical
  Journal, 862, 124, \dodoi{10.3847/1538-4357/aac9c4}

\bibitem[{Ruiz-Lapuente {et~al.}(2004)Ruiz-Lapuente, Comeron, Méndez, Canal,
  Smartt, Filippenko, Kurucz, Chornock, Foley, Stanishev, \&
  Ibata}]{ruiz-lapuente_binary_2004}
Ruiz-Lapuente, P., Comeron, F., Méndez, J., {et~al.} 2004, Nature, 431, 1069,
  \dodoi{10.1038/nature03006}

\bibitem[{Ruiz-Lapuente {et~al.}(2019)Ruiz-Lapuente, González~Hernández, Mor,
  Romero-Gómez, Miret-Roig, Figueras, Bedin, Canal, \&
  Méndez}]{ruiz-lapuente_tychos_2019}
Ruiz-Lapuente, P., González~Hernández, J.~I., Mor, R., {et~al.} 2019, The
  Astrophysical Journal, 870, 135, \dodoi{10.3847/1538-4357/aaf1c1}

\bibitem[{Sand {et~al.}(2016)Sand, Hsiao, Banerjee, Marion, Diamond, Joshi,
  Parrent, Phillips, Stritzinger, \& Venkataraman}]{sand_post-maximum_2016}
Sand, D.~J., Hsiao, E.~Y., Banerjee, D. P.~K., {et~al.} 2016, The Astrophysical
  Journal, 822, L16, \dodoi{10.3847/2041-8205/822/1/L16}

\bibitem[{Sand {et~al.}(2018)Sand, Graham, Botyánszki, Hiramatsu, McCully,
  Valenti, Hosseinzadeh, Howell, Burke, Cartier, Diamond, Hsiao, Jha, Kasen,
  Kumar, Marion, Suntzeff, Tartaglia, Wheeler, \& Wyatt}]{sand_nebular_2018}
Sand, D.~J., Graham, M.~L., Botyánszki, J., {et~al.} 2018, The Astrophysical
  Journal, 863, 24, \dodoi{10.3847/1538-4357/aacde8}

\bibitem[{Schaefer \& Pagnotta(2012)}]{schaefer_absence_2012}
Schaefer, B.~E., \& Pagnotta, A. 2012, Nature, 481, 164,
  \dodoi{10.1038/nature10692}

\bibitem[{Schechter {et~al.}(1993)Schechter, Mateo, \&
  Saha}]{schechter_dophot_1993}
Schechter, P.~L., Mateo, M., \& Saha, A. 1993, Publications of the Astronomical
  Society of the Pacific, 105, 1342, \dodoi{10.1086/133316}

\bibitem[{Schlafly \& Finkbeiner(2011)}]{schlafly_measuring_2011}
Schlafly, E.~F., \& Finkbeiner, D.~P. 2011, The Astrophysical Journal, 737,
  103, \dodoi{10.1088/0004-637X/737/2/103}

\bibitem[{Shappee {et~al.}(2013{\natexlab{a}})Shappee, Kochanek, \&
  Stanek}]{shappee_type_2013}
Shappee, B.~J., Kochanek, C.~S., \& Stanek, K.~Z. 2013{\natexlab{a}},
  {\textbackslash}apj, 765, 150, \dodoi{10.1088/0004-637X/765/2/150}

\bibitem[{Shappee {et~al.}(2018)Shappee, Piro, Stanek, Patel, Margutti,
  Lipunov, \& Pogge}]{shappee_strong_2018}
Shappee, B.~J., Piro, A.~L., Stanek, K.~Z., {et~al.} 2018, The Astrophysical
  Journal, 855, 6, \dodoi{10.3847/1538-4357/aaa1e9}

\bibitem[{Shappee {et~al.}(2013{\natexlab{b}})Shappee, Stanek, Pogge, \&
  Garnavich}]{shappee_no_2013}
Shappee, B.~J., Stanek, K.~Z., Pogge, R.~W., \& Garnavich, P.~M.
  2013{\natexlab{b}}, The Astrophysical Journal, 762, L5,
  \dodoi{10.1088/2041-8205/762/1/L5}

\bibitem[{Shappee {et~al.}(2016)Shappee, Piro, Holoien, Prieto, Contreras,
  Itagaki, Burns, Kochanek, Stanek, Alper, Basu, Beacom, Bersier, Brimacombe,
  Conseil, Danilet, Dong, Falco, Grupe, Hsiao, Kiyota, Morrell, Nicolas,
  Phillips, Pojmanski, Simonian, Stritzinger, Szczygieł, Taddia, Thompson,
  Thorstensen, Wagner, \& Woźniak}]{shappee_young_2016}
Shappee, B.~J., Piro, A.~L., Holoien, T. W.~S., {et~al.} 2016, The
  Astrophysical Journal, 826, 144, \dodoi{10.3847/0004-637X/826/2/144}

\bibitem[{Shen \& Bildsten(2014)}]{shen_ignition_2014}
Shen, K.~J., \& Bildsten, L. 2014, The Astrophysical Journal, 785, 61,
  \dodoi{10.1088/0004-637X/785/1/61}

\bibitem[{Shen \& Schwab(2017)}]{shen_wait_2017}
Shen, K.~J., \& Schwab, J. 2017, \dodoi{10.3847/1538-4357/834/2/180}

\bibitem[{Shen {et~al.}(2018)Shen, Boubert, Gänsicke, Jha, Andrews, Chomiuk,
  Foley, Fraser, Gromadzki, Guillochon, Kotze, Maguire, Siebert, Smith,
  Strader, Badenes, Kerzendorf, Koester, Kromer, Miles, Pakmor, Schwab, Toloza,
  Toonen, Townsley, \& Williams}]{shen_three_2018}
Shen, K.~J., Boubert, D., Gänsicke, B.~T., {et~al.} 2018, {\textbackslash}apj,
  865, 15, \dodoi{10.3847/1538-4357/aad55b}

\bibitem[{{The Astropy Collaboration} {et~al.}(2013){The Astropy
  Collaboration}, Robitaille, Tollerud, Greenfield, Droettboom, Bray, Aldcroft,
  Davis, Ginsburg, Price-Whelan, Kerzendorf, Conley, Crighton, Barbary, Muna,
  Ferguson, Grollier, Parikh, Nair, Günther, Deil, Woillez, Conseil, Kramer,
  Turner, Singer, Fox, Weaver, Zabalza, Edwards, Azalee~Bostroem, Burke, Casey,
  Crawford, Dencheva, Ely, Jenness, Labrie, Lim, Pierfederici, Pontzen, Ptak,
  Refsdal, Servillat, \& Streicher}]{the_astropy_collaboration_astropy_2013}
{The Astropy Collaboration}, Robitaille, T.~P., Tollerud, E.~J., {et~al.} 2013,
  Astronomy \& Astrophysics, 558, A33, \dodoi{10.1051/0004-6361/201322068}

\bibitem[{Timmes {et~al.}(1995)Timmes, Woosley, \&
  Weaver}]{timmes_galactic_1995}
Timmes, F.~X., Woosley, S.~E., \& Weaver, T.~A. 1995, The Astrophysical Journal
  Supplement Series, 98, 617, \dodoi{10.1086/192172}

\bibitem[{Tucker {et~al.}(2020)Tucker, Shappee, Vallely, Stanek, Prieto,
  Botyanszki, Kochanek, Anderson, Brown, Galbany, Holoien, Hsiao, Kumar,
  Kuncarayakti, Morrell, Phillips, Stritzinger, \&
  Thompson}]{tucker_nebular_2020}
Tucker, M.~A., Shappee, B.~J., Vallely, P.~J., {et~al.} 2020, Monthly Notices
  of the Royal Astronomical Society, 493, 1044, \dodoi{10.1093/mnras/stz3390}

\bibitem[{Valdes {et~al.}(2014)Valdes, Gruendl, \& {DES
  Project}}]{valdes_decam_2014}
Valdes, F., Gruendl, R., \& {DES Project}. 2014, 485, 379.
\newblock \url{https://ui.adsabs.harvard.edu/abs/2014ASPC..485..379V}

\bibitem[{Vallely {et~al.}(2019)Vallely, Fausnaugh, Jha, Tucker, Eweis,
  Shappee, Kochanek, Stanek, Chen, Dong, Prieto, Sukhbold, Thompson,
  Brimacombe, Stritzinger, Holoien, Buckley, Gromadzki, \&
  Bose}]{vallely_asassn-18tb_2019}
Vallely, P.~J., Fausnaugh, M., Jha, S.~W., {et~al.} 2019, Monthly Notices of
  the Royal Astronomical Society, 487, 2372, \dodoi{10.1093/mnras/stz1445}

\bibitem[{Virtanen {et~al.}(2020)Virtanen, Gommers, Oliphant, Haberland, Reddy,
  Cournapeau, Burovski, Peterson, Weckesser, Bright, van~der Walt, Brett,
  Wilson, Millman, Mayorov, Nelson, Jones, Kern, Larson, Carey, Polat, Feng,
  Moore, VanderPlas, Laxalde, Perktold, Cimrman, Henriksen, Quintero, Harris,
  Archibald, Ribeiro, Pedregosa, \& van Mulbregt}]{virtanen_scipy_2020}
Virtanen, P., Gommers, R., Oliphant, T.~E., {et~al.} 2020, Nature Methods, 17,
  261, \dodoi{10.1038/s41592-019-0686-2}

\bibitem[{Webbink(1984)}]{webbink_double_1984}
Webbink, R.~F. 1984, The Astrophysical Journal, 277, 355,
  \dodoi{10.1086/161701}

\bibitem[{Werner {et~al.}(2004)Werner, Roellig, Low, Rieke, Rieke, Hoffmann,
  Young, Houck, Brandl, Fazio, Hora, Gehrz, Helou, Soifer, Stauffer, Keene,
  Eisenhardt, Gallagher, Gautier, Irace, Lawrence, Simmons, Van~Cleve, Jura,
  Wright, \& Cruikshank}]{werner_spitzer_2004}
Werner, M.~W., Roellig, T.~L., Low, F.~J., {et~al.} 2004, The Astrophysical
  Journal Supplement Series, 154, 1, \dodoi{10.1086/422992}

\bibitem[{Whelan \& Iben(1973)}]{whelan_binaries_1973}
Whelan, J., \& Iben, Jr., I. 1973, The Astrophysical Journal, 186, 1007,
  \dodoi{10.1086/152565}

\bibitem[{Williams {et~al.}(2013)Williams, Borkowski, Ghavamian, Hewitt, Mao,
  Petre, Reynolds, \& Blondin}]{williams_azimuthal_2013}
Williams, B.~J., Borkowski, K.~J., Ghavamian, P., {et~al.} 2013, The
  Astrophysical Journal, 770, 129, \dodoi{10.1088/0004-637X/770/2/129}

\bibitem[{Winkler {et~al.}(2002)Winkler, Gupta, \& Long}]{winkler_sn_2002}
Winkler, P.~F., Gupta, G., \& Long, K.~S. 2002, \dodoi{10.1086/345985}

\bibitem[{Winkler {et~al.}(2005)Winkler, Long, Hamilton, \&
  Fesen}]{winkler_probing_2005}
Winkler, P.~F., Long, K.~S., Hamilton, A. J.~S., \& Fesen, R.~A. 2005, The
  Astrophysical Journal, 624, 189, \dodoi{10.1086/429155}

\bibitem[{Woods {et~al.}(2017)Woods, Ghavamian, Badenes, \&
  Gilfanov}]{woods_no_2017}
Woods, T.~E., Ghavamian, P., Badenes, C., \& Gilfanov, M. 2017, Nature
  Astronomy, 1, 800, \dodoi{10.1038/s41550-017-0263-5}

\bibitem[{Zheng {et~al.}(2013)Zheng, Silverman, Filippenko, Kasen, Nugent,
  Graham, Wang, Valenti, Ciabattari, Kelly, Fox, Shivvers, Clubb, Cenko, Balam,
  Howell, Hsiao, Li, Marion, Sand, Vinko, Wheeler, \& Zhang}]{zheng_very_2013}
Zheng, W., Silverman, J.~M., Filippenko, A.~V., {et~al.} 2013, The
  Astrophysical Journal, 778, L15, \dodoi{10.1088/2041-8205/778/1/L15}

\end{thebibliography}
\bibliographystyle{aasjournal}



\section{Contributor Roles}

\begin{itemize}
\item{Conceptualization: Joshua V. Shields, Wolfgang Kerzendorf, Ken Shen}
\item{Data curation: Joshua V. Shields, Armin Rest, Giovanni Strampelli}
\item{Formal Analysis: Joshua V. Shields, Armin Rest}
\item{Funding acquisition: Wolfgang Kerzendorf}
\item{Investigation: Joshua V. Shields, Wolfgang Kerzendorf, Alfredo Zenteno}
\item{Methodology: Joshua V. Shields, Matt Hosek, Armin Rest, Ken Shen, Tuan Do}
\item{Project administration: Wolfgang Kerzendorf}
\item{Resources: Armin Rest}
\item{Software: Jessica Lu, Matt Hosesk, Armin Rest}
\item{Supervision: Wolfgang Kerzendorf} 
\item{Validation: Joshua V. Shields}
\item{Visualization: Joshua V. Shields}
\item{Writing - original draft: Joshua V. Shields, Wolfgang Kerzendorf}
\item{Writing - review \& editing: Joshua V. Shields, Wolfgang Kerzendorf, Matt Hosek, Ken Shen, Tuan Do, Andrew Fullard}

\end{itemize}

\appendix \label{sec:appendix}

\section{Position Uncertainty Estimation} \label{App:pos}

We estimated the uncertainty of a single observation from the Dark Energy Camera following \cite{rest_cosmological_2013}:
\begin{equation}
    \sigma_{pix} =  0.1^2 + 1.5(\frac{FWHM}{SNR})^2
    \label{eq:uncertainty}
\end{equation}

\section{Matching Algorithm Choices} \label{App:match}

\begin{figure} [h!]
    \centering
    \includegraphics[width=.5\textwidth]{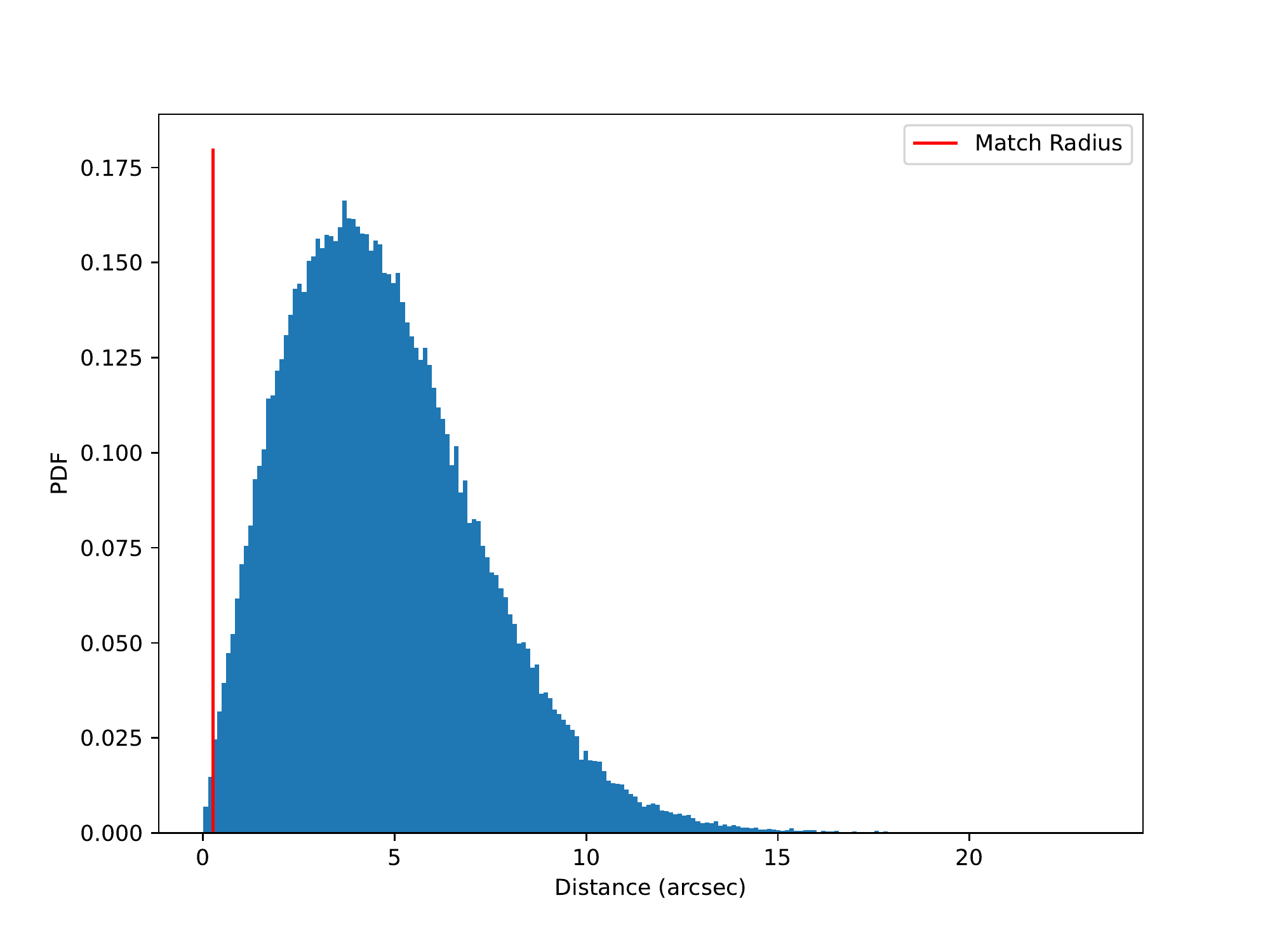}
    \caption{Nearest neighbor distribution function for a sample CCD after the initial stellar matching was completed. We adopted a search radius of 1 pixel or 0.263 arcsec for our initial matching algorithm, a small enough search radius to not allow for significant source confusion at the stellar density of the field. 
}
    \label{fig:sep_plot}
\end{figure}

To create our individual epoch catalogues we combined detections of sources with two or more observations using the \textsc{scipy.spatial.KDtree} package to subdivide the parameter space and reduce considered associations. We disregarded sources within three pixels from the edge of an observation, and only matched stars within one pixel or 0.263 arcec. We chose this search radius, taking in to consideration the stellar density of roughly one in eight pixels. Additionally, the nearest neighbor distribution function is shown in Fig \ref{fig:sep_plot}, which shows that the nearest neighbor of a source is farther than 0.263 arcsec away in almost all cases. A source could be matched with an unrelated object if it is both within the search region of the unrelated source and it is not detected in a given epoch. This likely has the effect of artificially increasing proper motions, as the source would be reported having moved anomalously far between epochs, and is much more likely to be an issue for faint objects. This effect may explain some of the faint, high proper motion objects in our sample.

\section{Proper Motion and Position Measurements Compared To Gaia} \label{App:compare}

\begin{figure}
    \centering
    \includegraphics[width=.45\textwidth]{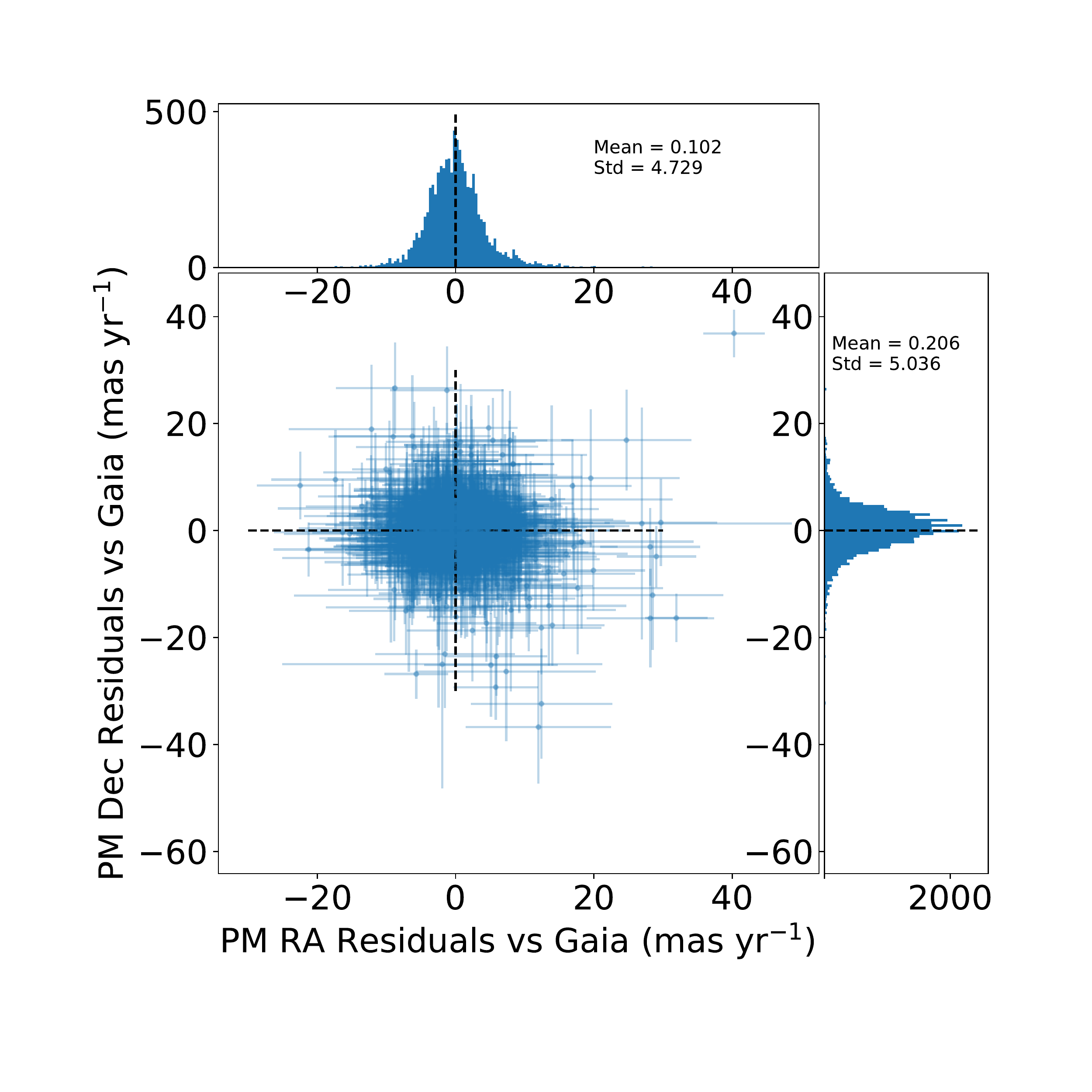}
    \includegraphics[width=.45\textwidth]{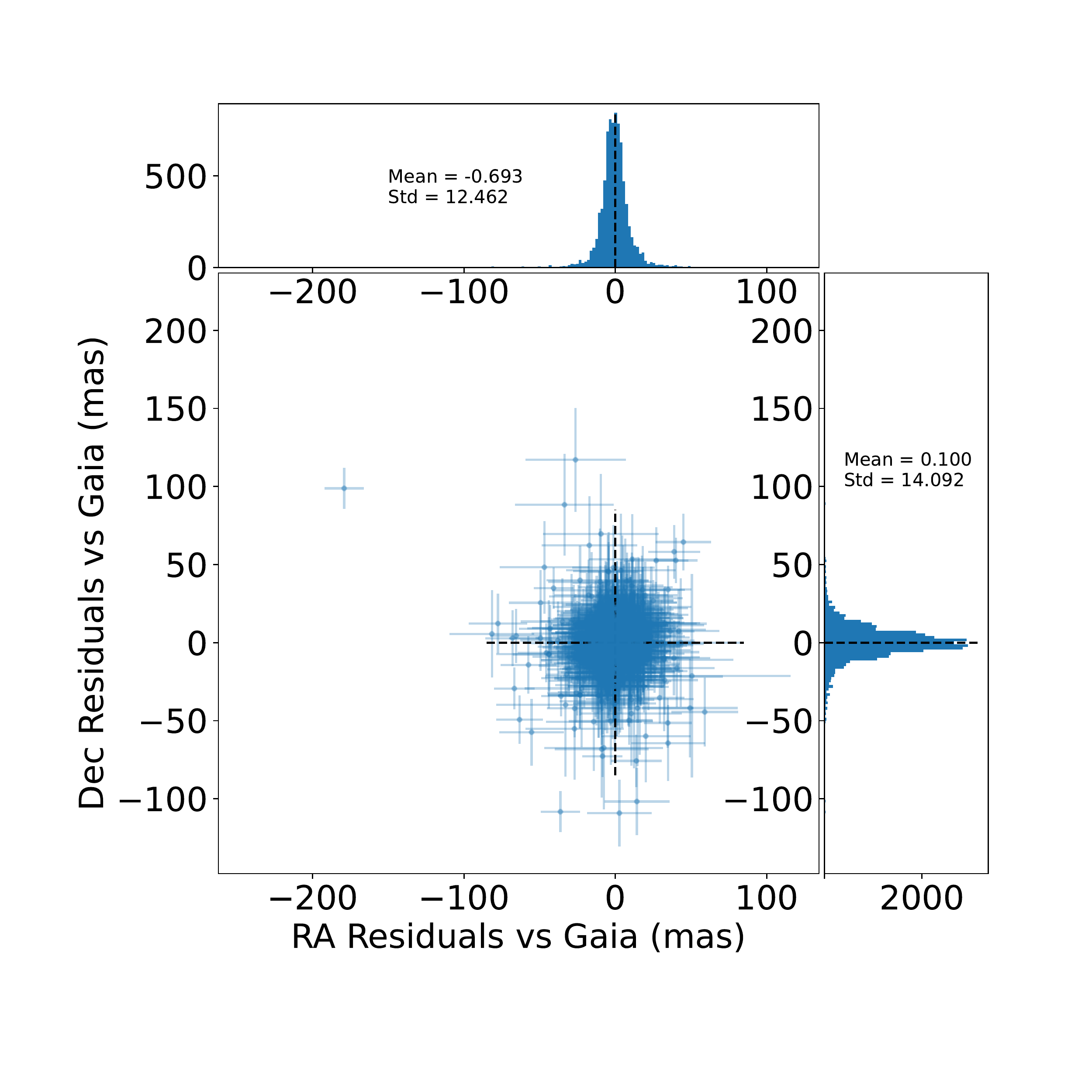}
    \caption{Residuals of DECam measurements vs. Gaia for cross-matched sources in RA and Dec in both proper motion (left) and position (right) space. Our desired precision was 80 mas yr$^{-1}$, many times higher than the scatter in our distributions which give an empirical estimate of our error. 
    }
    \label{fig:residuals}
\end{figure}

\begin{figure}
    \centering
    \includegraphics[width=.45\textwidth]{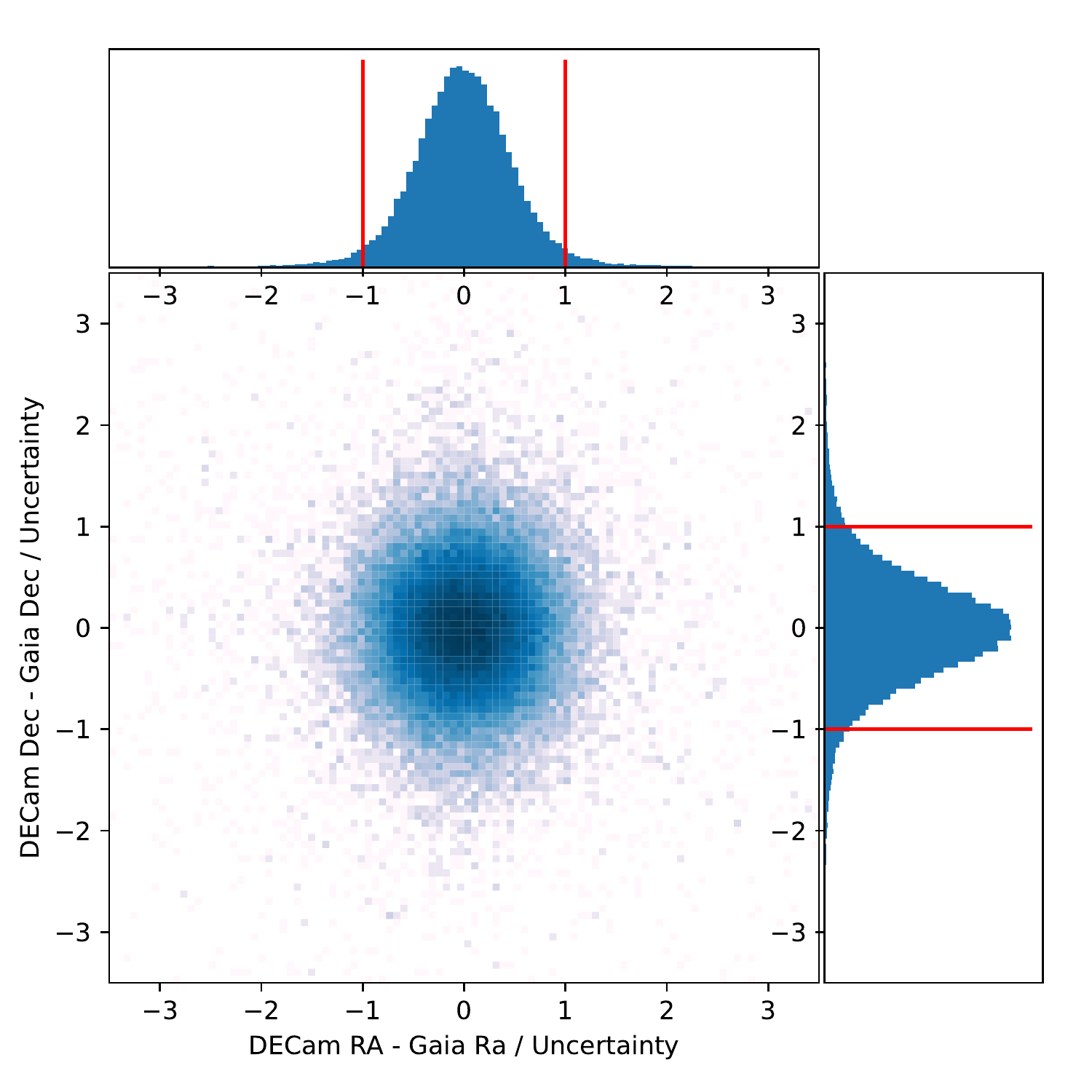}
    \includegraphics[width=.45\textwidth]{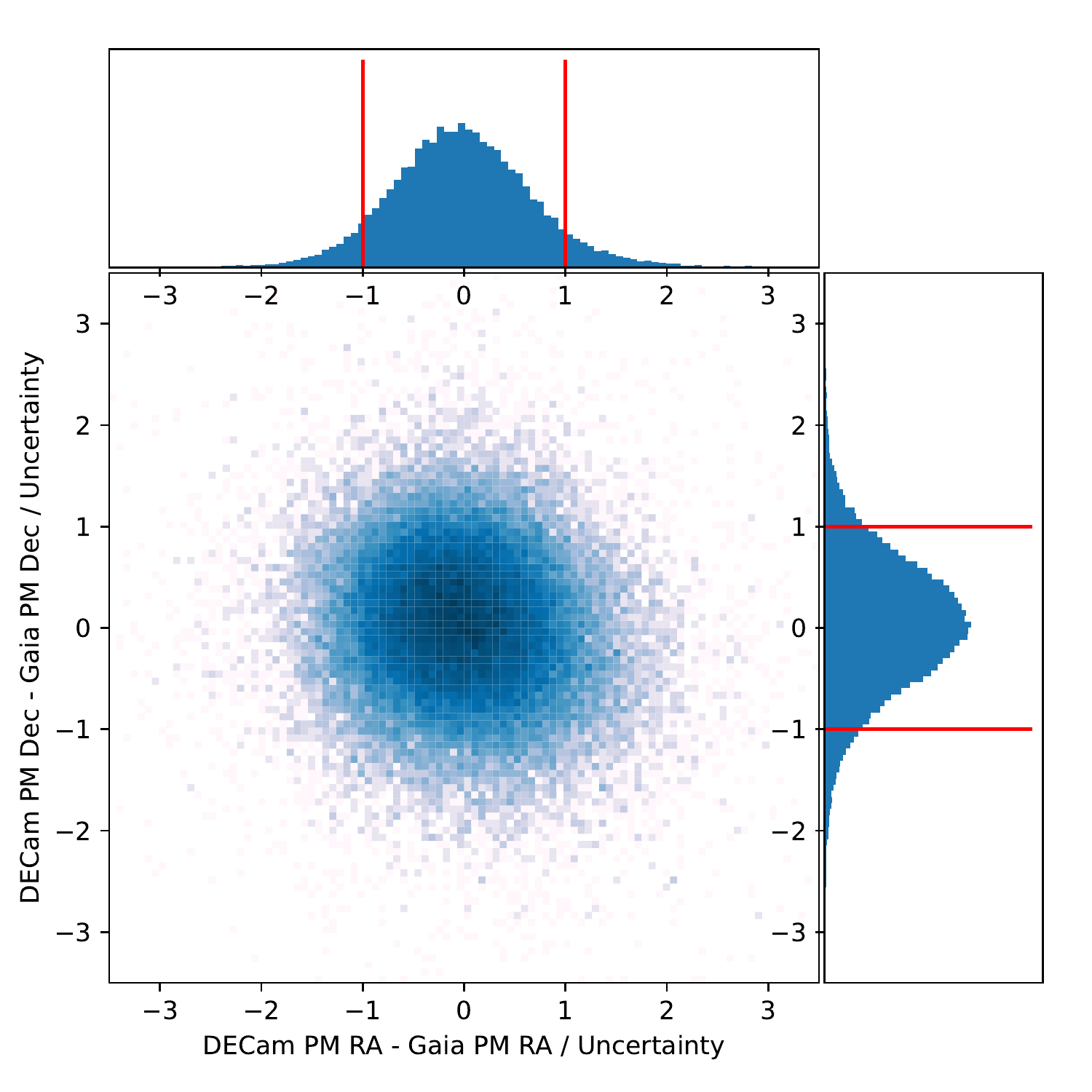}
    \caption{Position (left) and proper motion (right) residuals of DECam vs. Gaia measurements for cross-matched sources in RA and Dec over uncertainty in each dimension. We expect $68\%$ of sources to be smaller than 1 sigma. We find $\sim92\%$ of sources in position space and $\sim85\%$ of sources in proper motion space lie within this region, suggesting that our errors may be overestimated. }
    \label{fig:diff_err}
\end{figure}

We compare the proper motions and positions of sources matched between the DECam and Gaia EDR3 catalogues. Our residual proper motion and position measurements are consistently smaller than our desired signal of 80 mas yr$^{-1}$ (Figures \ref{fig:residuals} and \ref{fig:diff_err}), verifying the accuracy of the astrometric reference frame for our DECam measurements. We note that these matched stars were previously used to discover the polynomial transformation, so these do not provide a completely independent verification, but the fitting free parameters (30) are three orders of magnitude fewer than the sampled data-points ($\sim$ 8000 per CCD) so the correlations are not be severe.

\end{document}